\documentstyle[12pt,aaspp4,flushrt]{article}
\begin{document}
\def\gtorder{\mathrel{\raise.3ex\hbox{$>$}\mkern-14mu
             \lower0.6ex\hbox{$\sim$}}}
\def\ltorder{\mathrel{\raise.3ex\hbox{$<$}\mkern-14mu
             \lower0.6ex\hbox{$\sim$}}}

\title{Clusters and Superclusters of Galaxies} 
\author{Neta A. Bahcall\\Princeton University
Observatory\\
Princeton, NJ 08544}

\begin{abstract}
Rich clusters of galaxies are the most massive virialized
systems known.  Even though they contain only a small fraction of all
galaxies, rich clusters provide a powerful tool for the study of
galaxy formation, dark matter, large-scale structure, and cosmology.

Superclusters, the largest known systems of galaxies, extend to
$\sim 100h^{-1}$ Mpc in size and highlight the large-scale structure of
the universe.  This large-scale structure reflects initial conditions in the early
universe and places strong contraints on models of galaxy formation and
on cosmology.

Some of the questions that can be addressed with clusters and
superclusters of galaxies include: How did galaxies and larger
structures form and evolve? What is the amount, composition, and
distribution of matter in clusters and larger structures? How does
the cluster mass density relate to the matter density 
in the universe? What constraints
can the cluster and supercluster data place on cosmology?

I will discuss in these lectures some of the properties of clusters
and superclusters of galaxies that can be used to investigate these topics.
\end{abstract}

\section{Introduction}
\label{sec:1}

Clusters and superclusters of galaxies have been studied extensively
both 
for their intrinsic properties and to investigate the dark matter in the
universe, the baryon content of the universe, large-scale structure, evolution, and
cosmology.  For previous reviews see Zwicky (1958), Bahcall (1977,
1988, 1996), Oort (1983), Dressler (1984), Rood (1988), and Peebles
(1993).

In these lectures I discuss the following topics and their
implications for structure formation and cosmology.

\begin{description}
\item [\hfil Section 2:] Optical properties of galaxy clusters
\item [\hfil Section 3:] X-ray properties of galaxy clusters
\item [\hfil Section 4:] The baryon fraction in clusters
\item [\hfil Section 5:] Cluster masses
\item [\hfil Section 6:] Where is the dark matter?
\item [\hfil Section 7:] The mass function of clusters
\item [\hfil Section 8:] Quasar-cluster association
\item [\hfil Section 9:] Superclusters
\item [Section 10:] The cluster correlation function
\item [Section 11:] Peculiar motions of clusters
\item [Section 12:] Some unsolved problems
\end{description}
A Hubble constant of $H_o = 100h~{\rm km~s^{-1}~Mpc^{-1}}$ is used
throughout.  

\section{Optical Properties of Galaxy Clusters}
\label{sec:2}

\subsection{Typical Properties of Clusters and Groups}
\label{sec:2.1}

Clusters of galaxies are bound, virialized, high overdensity systems
of galaxies, held together by the clusters self gravity.  Rich
clusters contain, by traditional definition (Abell 1958), at least 30
galaxies brighter than $m_3 + 2^m$ (where $m_3$ is the magnitude of the
third brightest cluster member) within a radius of $R \simeq
1.5h^{-1}~{\rm Mpc}$ of the cluster center.  This galaxy count is
generally defined as the richness of the cluster.  The galaxies in
rich clusters move with random peculiar velocities of typically $\sim
750~{\rm km~s^{-1}}$ (median line-of-sight velocity dispersion).  This
motion corresponds to a typical rich cluster mass (within
$1.5h^{-1}~{\rm Mpc}$) of $\sim 5 \times 10^{14}h^{-1}~M_\odot$.  In
addition to galaxies, all rich clusters contain an intracluster medium
of hot plasma, extending as far as the main galaxy concentration ($R
\sim 1.5h^{-1}~{\rm Mpc}$).  The typical temperature of the hot
intracluster gas is $\sim 5$~kev, with a range from $\sim 2$ to
14~kev; the central gas density is $\sim 10^{-3}$ electrons ${\rm
cm^{-3}}$.  The hot plasma is detected through the luminous X-ray
emission it produces by thermal bremsstrahlung radiation, with $L_x
\sim 10^{44}~{\rm erg\ s^{-1}}$.  

Like mountain peaks on earth, the
high density rich clusters are relatively ``rare'' objects; they
exhibit a spatial number density of $\sim 10^{-5}$ clusters ${\rm
Mpc^{-3}}$, as compared with $\sim 10^{-2}$ galaxies ${\rm Mpc^{-3}}$ for
the density of bright galaxies.

\begin{table}
\begin{minipage}{5in}
\caption{Typical Properties of Clusters and Groups.\protect\label{tab:one}}
\begin{tabular*}{5in}{lll}
\noalign{\medskip\hrule\smallskip\hrule\medskip}
\multicolumn{1}{c}{Property$^{a}$}  & \multicolumn{1}{c}{Rich
clusters} 
& \multicolumn{1}{c}{Groups\ \ \ } \\
&&\multicolumn{1}{c}{and poor clusters}\\
\noalign{\medskip\hrule\medskip}
Richness$^{b}$ & 30--300 galaxies &  3--30 galaxies \\[5pt]
Radius$^{c}$ & (1--2) $h^{-1}$ Mpc & (0.1--1) $h^{-1}$ Mpc \\[5pt]
Radial velocity dispersion$^{d}$ & 400--1400 km$\,{\rm s}^{-1}$ 
& 100--500 km$\,{\rm s}^{-1}$ \\[5pt]
Radial velocity dispersion$^d$& $\sim 750$ km$\,{\rm s}^{-1}$ 
& $\sim 250$ km$\,{\rm s}^{\rm -1}$\\
 (median)\\ [5pt]
Mass $(r \leq 1.5h^{-1}$ Mpc)$^e$&(10$^{14 }$--2$\times 10^{15})h^{-1}$ 
${\cal M}_\odot$ & (10$^{12.5}$--10$^{14})h^{-1}$ ${\cal M}_\odot$ \\[5pt]
Luminosity $(B)$$^f$ & $(6\times 10^{\rm 11 }$--6$\times 
10^{\rm 12})h^{\rm -2}$ $L_\odot$ & 
(10$^{\rm 10.5}$--10$^{\rm 12})h^{\rm -2}$ $L_\odot$\\
$(r \leq  1.5 h^{\rm -1}$ Mpc) && \\[5pt]
$\langle {\cal M}/L_{\rm B}\rangle$$^g$ & $\sim 300h$ 
${\cal M}_\odot /L_\odot$ & $\sim 200h$ ${\cal M}_\odot /L_\odot$\\[5pt]
X-ray temperature$^h$ & 2--14 keV & $\ltorder 2$ keV \\[5pt]
X-ray luminosity$^h$ & (10$^{\rm 42.5}$--10$^{\rm 45})h^{\rm -2}$ 
${\rm erg\,s}^{\rm -1}$ & $\ltorder 10^{\rm 43}h^{\rm -2}$ 
${\rm erg\,s}^{\rm -1}$ \\[5pt]
Cluster number density$^i$ & (10$^{\rm -5}$--10$^{-6})h^{\rm 3}$ Mpc$^{\rm
-3}$ 
&  (10$^{\rm -3}$--10$^{\rm -5})h^{\rm 3}$ Mpc$^{\rm -3}$\\[5pt]
Cluster correlation scale$^j$ & $(22\pm 4)h^{\rm -1}$ Mpc 
$(R \geq 1)$ & $(13\pm 2)h^{\rm -1}$ Mpc\\[5pt]
Fraction of galaxies in & $\sim 5\%$ &$\sim 55\%$\\
clusters or groups$^k$ && \\
\noalign{\medskip\hrule\medskip}
\end{tabular*}
\begin{itemize}
\item[$^a$]In most entries, the typical range in the listed property or
the median value is given.  Groups and poor clusters are a
natural and continuous extension to lower richness, mass, size,
and luminosity from the rich and rare clusters.
\item[$^b$]Cluster richness:  the number of cluster galaxies
brighter than $m_{\rm 3}$ + 2$^{m}$ (where $m_{\rm 3}$ is the magnitude of the third
brightest cluster galaxy), and located within a 1.5$h^{\rm -1}$ Mpc
radius of the cluster center (\S\ref{sec:2.2}).
\item[$^c$]The radius of the main concentration of galaxies (where,
typically, the galaxy surface density drops to $\sim 1\%$ of the
central 
core density).  Many clusters and groups are embedded in larger
scale structures (to tens of Mpc).
\item[$^d$]Typical observed range and median value for the radial
(line-of-sight) velocity dispersion in groups and clusters (\S\ref{sec:2.9}).
\item[$^e$]Dynamical mass range of clusters within 1.5$h^{\rm -1}$ Mpc
radius (\S\ref{sec:2.10}).
\item[$^f$]Luminosity range (blue) of clusters within 1.5$h^{\rm -1}$ Mpc
radius (\S\ref{sec:2.10}).
\item[$^g$]Typical mass-to-light ratio of clusters and groups (median
value) (\S\ref{sec:2.10}).
\item[$^h$]Typical observed ranges of the X-ray temperature and 2--10-keV
X-ray luminosity of the hot intracluster gas (\S\ref{sec:3}).
\item[$^i$]The number density of clusters decreases sharply with cluster
richness (\S\ref{sec:7}).
\item[$^j$]The cluster correlation scale for rich $(R\geq 1$, $N_{R}\geq
50$, $n_{c}=0.6 \times 10^{\rm -5}$$h^{\rm 3}$ Mpc$^{\rm -3}$) and poor $(N_{R}
\gtorder 20$, $n_{c}=2.4 \times 10^{\rm -5}$
$h^{\rm 3}$ Mpc$^{\rm -3}$) clusters (\S\ref{sec:2.3}, \S\ref{sec:7}).
\item[$^k$]The fraction of bright galaxies $(\gtorder L^{\rm *}$) 
in clusters and groups
within $1.5 h^{\rm -1}$ Mpc.
\end{itemize}
\end{minipage}
\end{table}

The main properties of clusters and groups of galaxies are summarized
in Table~\ref{tab:one} (Bahcall 1996).  The table lists the typical
range and/or median value of each observed property.  Groups and poor
clusters, whose properties are also listed, provide a natural and
continuous extension to lower richness, mass, size, and luminosity
from the rich and rare clusters.

In the following subsections I discuss in more detail some of these
intrinsic cluster properties.

\subsection{Distribution of Clusters with Richness and Distance}
\label{sec:2.2}

\begin{table}
\begin{minipage}{4in}
\caption{Distribution of Abell Clusters with Distance and
Richness.$^a$\protect\label{tab:two}}
\begin{tabular}{c@{\extracolsep{\fill}}rrclr}
\noalign{\medskip\hrule\smallskip\hrule\medskip}
\multicolumn{3}{c}{ Distance distribution} 
& \multicolumn{3}{c}{Richness distribution} \\
$D$ & $\langle z_{\rm est}\rangle$ & $N_{\rm cl}(R\geq 1)$ 
& $R$ & $N_{R}$ & $N_{\rm cl}$\\
\noalign{\medskip\hrule\medskip}
1 & 0.0283 & 9 & (0)$^{b}$ & (30--49) & $(\sim 10^{\rm 3})$\\
2 & 0.0400 & 2 & 1 & 50--79 & 1224\\
3 & 0.0577 & 33 & 2 & 80--129 & 383\\
4 & 0.0787 & 60 & 3 & 130--199 & 68\\
5 & 0.131 & 657 & 4 & 200--299 & 6\\
6 & 0.198 & 921 & 5 & $\geq 300$ & 1 \\
&\multicolumn{1}{r}{Total} &  $\overline{1682}$ 
&& \multicolumn{1}{r}{Total $(R\geq 1)$} &  $\overline{1682}$\\
\noalign{\medskip\hrule\medskip}
\multicolumn{3}{c}{Nearby redshift sample$^{c,d}$} & \multicolumn{3}{c}{Distant projected sample$^{d}$}\\
\multicolumn{3}{c}{$D\leq 4$}  & & \multicolumn{1}{c}{$D=5+6$} & \\
\noalign{\medskip\hrule\medskip}
\multicolumn{2}{l}{$N_{\rm cl}$(total)} & 104 & &\multicolumn{1}{c}{1574}& \\
\multicolumn{2}{l}{$N_{\rm cl}(b \geq 30^{\circ})$} & 71 & 
&\multicolumn{1}{c}{984} & \\
\multicolumn{2}{l}{$N_{\rm cl}(b\leq -30^{\circ})$} & 33 &
&\multicolumn{1}{c}{563} & \\
\multicolumn{2}{l}{$N_{\rm cl}(R=1)$} & 82 & &\multicolumn{1}{c}{1125} & \\
\multicolumn{2}{l}{$N_{\rm cl}(R\geq 2)$} & 22 & &\multicolumn{1}{c}{422} & \\
\noalign{\medskip\hrule\medskip}
\end{tabular}
\begin{itemize}
\item[$^a$]Statistical sample.  $|b|$ boundaries as given in 
Table 1 of Abell (1958).
Notation:  $D=$ distance group (defined by the estimated redshifts of
the clusters);
$\langle z_{\rm est}\rangle =
$ average estimated redshift from the
magnitude of the tenth brightest galaxy; $N_{\rm cl} =$ number of
clusters;
$R=$ richness class; $N_{R} =$ number of galaxies brighter than
$m_{3}+2^{m}$ within $R_{A}=1.5h^{\rm -1}$ Mpc (richness count).
\item[$^b$]$R=0$ clusters are not part of the statistical sample.
\item[$^c$]Redshifts by Hoessel et~al. (1980).
\item[$^d$]This sample is limited to $|b|\geq 30^{\circ}$ in addition to the
$|b|$ boundaries of the statistical sample.
\end{itemize}
\end{minipage}
\end{table}

An illustration of the distribution of rich clusters with richness and
distance is summarized in Table~\ref{tab:two}.  The table refers to
the statistical sample of the Abell (1958) cluster catalog (i.e.,
richness class $R \geq 1$, corresponding to a richness threshold count
of $N_R \geq 50$ galaxies within $1.5h^{-1}$~Mpc radius and magnitude $m
\leq m_3 + 2^m$ (see Table~\ref{tab:one}); redshift range $z \simeq
0.02$ to 0.2; and sky coverage $\delta > -27^\circ$ and $|b| \gtorder
30^\circ$).  The Abell catalog covers $\sim 1/3$ of the entire sky to
$z \ltorder 0.2$.  Recent smaller automated surveys from digitized plates are
reported by the Edinburgh-Durham catalog (EDCC; Lumsden et al. 1992),
and the APM survey (Dalton et al. 1992).  A large automated cluster
catalog will be available in the near future from the Sloan Digital
Sky Survey (SDSS).  This and other planned surveys will allow a more
accurate determination of the distribution of rich clusters with
richness and distance and other statistical studies of clusters.

\subsection{Number Density of Clusters}
\label{sec:2.3}

The number density of clusters is a strong function of cluster
richness.  Integrated cluster densities, $n_{c}$ ($>N_{R}$), representing the number
density of clusters above a given richness threshold, and
the associated mean cluster separation, $d$ ($\equiv n_{c}^{\rm -1/3}$),
are listed in Table~\ref{tab:three} (Bahcall and Cen 1993).

\subsection{Fraction of Galaxies in Clusters}
\label{sec:2.4}

The fraction of galaxies in rich $R\gtorder 0$ clusters is  $\sim 5\%$
(within the Abell radius $R_{A} = 1.5 h^{\rm -1}$ Mpc).
The fraction of all galaxies that belong in clusters increases
with increasing radius $R_{A}$ and with decreasing cluster richness
threshold.

The average number of galaxies per cluster for $R\geq 0$ clusters
within $1.5h^{-1}$~Mpc radius and $m \leq m_3 + 2^m$ is $\langle
N_R\rangle_{\rm median} \simeq 50$, or $\langle N_R\rangle_{\rm mean}
\simeq 56$.  For $R \geq 1$ clusters the average is
$\langle N_R\rangle_{\rm median} \simeq 60$, or $\langle N_R\rangle_{\rm
mean} \simeq 75$.  The number of galaxies increases to fainter
luminosities following the Schechter (1976) luminosity function.

\subsection{Galaxy Overdensity in Rich Clusters}
\label{sec:2.5}

\begin{table}
\begin{minipage}{4.5in}
\caption[]{Number Density of Clusters.\protect\label{tab:three}}
\begin{tabular*}{4.5in}[ht]{c@{\extracolsep{\fill}}ccc}
\noalign{\medskip\hrule\smallskip\hrule\medskip}
$R$ & $N_{R}$ &  \multicolumn{1}{c}{$n_{c}$ ($>N_{R})h^{\rm 3}$ 
(Mpc$^{\rm -3})^a$} & $d$ ($> N_{R}) h^{\rm -1}$ (Mpc) \\
\noalign{\medskip\hrule\medskip}
$\geq 0$ & $\geq 30$ & \hspace{2.5pc}13.5 $\times 10^{\rm -6}$ & ~42\\
$\geq 1$ & $\geq 50$ & \hspace{2.5pc}6.0 $\times 10^{\rm -6}$ & ~55\\
$\geq 2$ & $\geq 80$ & \hspace{2.5pc}1.2 $\times 10^{\rm -6}$ & ~94\\
$\geq 3$ & $\geq 130$ & \hspace{2.5pc}1.5 $\times 10^{\rm -7}$ & 188\\
\noalign{\medskip\hrule}
\end{tabular*}
\begin{itemize}
\item[$^a$]Approximate uncertainties are 10$^{\rm \pm 0.2}$ for the $R\geq 0,1,2$
densities and 10$^{\pm 0.3}$ for $R\geq 3$.
\end{itemize}
\end{minipage}
\end{table}

The average number density of bright $(\gtorder L^{\rm *)}$ galaxies in
$R\gtorder 0$ clusters (within $R_{A} = 1.5 h^{\rm -1}$ Mpc) is
\begin{equation}
n_{g}{\rm (cluster)}\sim 3 h^{\rm 3}~{\rm galaxies~Mpc}^{\rm -3}.
\label{eq:1}
\end{equation}
The average overall (field) number density of bright $(\gtorder L^{\rm *})$
galaxies is
\begin{equation}
n_{g}{\rm (field)} \sim 1.5 \times 10^{\rm -2} h^{\rm 3}
~{\rm galaxies~Mpc}^{\rm -3}.
\label{eq:2}
\end{equation}
The average galaxy overdensity in rich $(R\geq 0)$ clusters (within
$1.5h^{-1}$~Mpc radius) is thus
\begin{equation}
n_{g}{\rm (cluster)} /n_{g}{\rm (field)} \sim 200.
\label{eq:3}
\end{equation}
The typical threshold galaxy overdensity in clusters (within
$1.5h^{-1}$~Mpc radius) is
\begin{equation}
R\geq 0~{\rm clusters:}~~n_{g}{\rm (cluster)} /n_{g}{\rm (field)}\gtorder 100,
\label{eq:4}
\end{equation}
\begin{equation}
R\geq 1~{\rm clusters:}~~n_{g}{\rm (cluster)}/n_{g}{\rm (field)}\gtorder 200.
\label{eq:5}
\end{equation}
The galaxy overdensity increases at smaller radii from the cluster
center. The galaxy overdensity in the cores of typical compact rich
clusters is approximately
\begin{equation}
n^{0}_{g}{\rm (cluster~core)}/n_{g}{\rm (field)}\sim 10^{\rm 4}{\rm -}10^5.
\label{eq:6}
\end{equation}

\subsection{Density Profile}
\label{sec:2.6}

The radial density distribution of galaxies in a rich cluster
can be approximated by a bounded Emden isothermal profile
 (Zwicky 1957; Bahcall 1977), or by its King approximation (King 1972)
in the {\it central} regions.

In the central regions, the King approximation for the galaxy
distribution is
\begin{equation}
n_{g}(r)=n^{0}_{g}(1+r^{\rm 2}/R_{c}^{\rm 2})^{\rm -3/2},
~~{\rm spatial~profile}
\label{eq:7}
\end{equation}
\begin{equation}
S_{g}(r)=S^{0}_{g}(1+r^{\rm 2}/R_{c}^{\rm 2})^{\rm -1  },
~~{\rm projected~profile}.
\label{eq:8}
\end{equation}
$n_{g}(r)$ and $S_g(r)$ are, respectively, the space and projected
profiles (of the number density of galaxies, or brightness), $n^{0}_g$
and $S^{0}_{g}$ are the respective central densities, and $R_{c}$ is the
cluster core radius [where $S(R_{c})=S^{0}/2]$.  Typical central
densities and core radii of clusters are listed in the following subsection.
The projected and space central densities relate as
\begin{equation}
S^{0}_{g}=2R_{c} n^{0}_{g}.
\label{eq:9}
\end{equation}
 A bounded Emden isothermal profile of galaxies in clusters
 yields a profile slope that varies approximately
as (Bahcall 1977)
\begin{equation}
S_{g}(r\ltorder R_{c}/3)\sim {\rm constant},
\label{eq:10}
\end{equation}
\begin{equation}
S_{g}(R_{c}\ltorder r\ltorder 10 R_{c}) \propto r ^{\rm -1.6 };
\label{eq:11}
\end{equation}
therefore
\begin{equation}
n_{g}(R_{c }\ltorder r\ltorder 10 R_{c}) \propto r^{\rm -2.6 }.
\label{eq:12}
\end{equation}
The galaxy--cluster cross-correlation function (Peebles 1980; Lilje
and Efstathiou 1988) also 
represents the average radial density distribution of galaxies
around clusters.  For $R\geq 1$ clusters, and $r$ in $h^{\rm -1}$ Mpc,
these references suggest, respectively
\begin{equation}
\xi_{gc}(r)\simeq 130r^{\rm -2.5}+70r^{\rm -1.7}
\label{eq:13}
\end{equation}
or
\begin{equation}
\xi_{gc}(r)\simeq 120r^{\rm -2.2}\ .
\label{eq:14}
\end{equation}
The average density distribution profile of galaxies in clusters
thus follows, approximately,
\begin{equation}
n_{g}(r)\propto r^{\rm -2.4 \pm  0.2}~~{\rm (spatial)},\ \ \ r > R_c
\label{eq:15}
\end{equation}
\begin{equation}
S_{g}(r)\propto r^{-1.4 \pm 0.2}~~{\rm (projected)},\ \ \ r > R_c\ .
\label{eq:16}
\end{equation}
Some substructure (subclumping) in the distribution of galaxies
 exists in a significant fraction of rich clusters ($\sim 40$\%)
 (Geller 1990).

\subsection{Central Density and Core Size}
\label{sec:2.7}

Central number density of galaxies in rich compact clusters
 (Bahcall 1975, 1977; Dressler 1978) is (for galaxies in 
the brightest 3 magnitude range)
\begin{equation}
n^{0}_{g}(\Delta m\simeq 3^{m})\sim 10^{\rm 3} h^{\rm 3}
~{\rm galaxies~Mpc}^{\rm -3}.
\label{eq:17}
\end{equation}
The central density reaches $\sim 10^{\rm 4} h^{\rm 3}$ galaxies$\,$Mpc$^{\rm -3}$
for the richest compact clusters.
The typical central {\it mass} density in rich compact clusters,
determined from cluster dynamics is
\begin{eqnarray}
\rho_{0}({\rm mass})&\simeq& 9\sigma_{r,c}^2/4\pi GR_c^2
\\
&\sim& 4 \times 10^{15} {\cal M}_{\odot}~{\rm Mpc}^{\rm -3}
[(\sigma_{r,c}/10^{\rm 3}~{\rm km\,s}^{\rm -1})/(R_{c}/0.2
~{\rm Mpc})]^{\rm 2} h^{\rm 2}\nonumber 
\label{eq:18}
\end{eqnarray}
where $\sigma_{r,c}$ is the radial (line-of-sight) central cluster velocity
dispersion (in km$\,{\rm s}^{\rm -1}$) and $R_{c}$ is the cluster core radius
(in Mpc).

Core radii of typical rich compact clusters, determined from the
galaxy distribution (Bahcall 1975; Dressler 1978; Sarazin 1986) are in
the range
\begin{equation}
R_c \simeq (0.1 - 0.25)h^{-1}~{\rm Mpc}
\label{eq:19}
\end{equation}
Core radii of the X-ray emitting intracluster gas (\S\ref{sec:3})
are
\begin{equation}
R_c ({\rm X-rays}) \simeq (0.1 - 0.3)h^{-1}~{\rm Mpc}
\label{eq:20}
\end{equation}
The core radius of the mass distribution determined from gravitational
lensing observations of some clusters may be smaller, $R_c \ltorder
50$~kpc, than determined by the galaxy and gas distribution.

The typical central density of the hot intracluster gas in rich
clusters (\S\ref{sec:3}) is
\begin{equation}
n_{e}\sim 10^{\rm -3}~{\rm electrons\,cm}^{\rm -3}.
\label{eq:21}
\end{equation}

\subsection{Galactic Content in Rich Clusters}
\label{sec:2.8}

The fraction of elliptical, S0, and spiral galaxies in rich
clusters differs from that in the field, and depends on the
classification type, or density, of the cluster (see \S\ref{sec:2.11})
 (Bahcall 1977; Oemler 1974; Dressler 1980).  See Table~\ref{tab:four}.

The fraction of elliptical (E) and S0 galaxies increases and the fraction
of spirals decreases toward the central cores of rich compact
clusters.  The fraction of spiral galaxies in the dense cores of
some rich clusters (e.g., the Coma cluster) may be close to zero.

The galactic content of clusters as represented in Table~\ref{tab:four}
is part of the general density--morphology relation of galaxies
 (Dressler 1980; Postman and Geller 1984); as the local 
density of galaxies increases, the fraction
of E and S0 galaxies increases and the fraction of spirals decreases.  For
local galaxy densities $n_{g}\ltorder 5$ galaxies Mpc$^{-3}$, the fractions remain
approximately constant at the average ``Field" fractions listed
above.

\begin{table}
\caption[]{Typical Galactic Content of 
Clusters $(r \ltorder 1.5 h^{-1}~{\rm Mpc})$.\protect\label{tab:four}}
\begin{tabular}{lcccc}
\noalign{\medskip\hrule\smallskip\hrule\medskip}
\multicolumn{1}{c}{Cluster type} & E & S0 & Sp & (E+S0)/Sp\\
\noalign{\medskip\hrule\medskip}
Regular clusters (cD) &   35\% & 45\% & 20\% & 4.0\\
Intermediate clusters (spiral-poor) & 20\% & 50\% & 30\% & 2.3\\
Irregular clusters (spiral-rich) & 15\% & 35\% & 50\% & 1.0\\
Field & 10\% & 20\% & 70\% & 0.5\\
\noalign{\medskip\hrule}
\end{tabular}
\end{table}

\subsection{Velocity Dispersion}
\label{sec:2.9}

The typical radial (line-of-sight) velocity dispersion of galaxies
in rich clusters (median value) is
\begin{equation}
\sigma_{r}\sim 750~{\rm km\,s}^{-1}.
\label{eq:22}
\end{equation}
The typical range of radial velocity dispersion in rich clusters
 (Struble and Rood 1991) is
\begin{equation}
\sigma_{r}\sim 400\mbox{--}1400~{\rm km\,s}^{-1}.
\label{eq:23}
\end{equation}
 A weak correlation between $\sigma_{r}$ and richness exists; richer
clusters exhibit, on average, larger velocity dispersion (Bahcall 1981).
 The observed velocity dispersion of galaxies in rich clusters is
generally consistent with the velocity implied by the X-ray
temperature of the hot intracluster gas (\S\ref{sec:3.5}), as well as
with the cluster velocity dispersion implied from observations of
gravitational lensing by clusters (except possibly in the central
core).
 Velocity dispersion and temperature profiles as a function of
distance from the cluster center have been measured only for a
small number of clusters so far.  The profiles are typically isothermal
$[\sigma^2_{r}(r)\sim T_{\rm x}(r)\sim {\rm constant}]$ for $r\ltorder
{\rm 0.5-1}h^{-1}$ Mpc,
and drop somewhat at larger distances.

\subsection{Mass, Luminosity, and Mass-to-Luminosity Ratio}
\label{sec:2.10}

The typical dynamical mass of rich clusters within $1.5 h^{-1}$ Mpc
radius sphere (determined from the virial theorem for an isothermal
distribution) is
\begin{eqnarray}
{\cal M}_{\rm cl}(\leq 1.5)\simeq {2\sigma_r^2(1.5h^{-1}~{\rm Mpc})\over G}
\simeq 0.7\times 10^{15}\Biggl({\sigma_r \over 1000}\Biggr)^2
\\
\simeq 0.4 \times 10^{15} h^{-1}{\cal M}_\odot
\ \,({\rm for}~\sigma_r\sim 750~{\rm km\,s}^{-1}).\nonumber 
\label{eq:24}
\end{eqnarray}
The approximate range of masses for $R\gtorder 0$ clusters (within
1.5$h^{-1}$ Mpc) is
\begin{equation}
{\cal M}_{\rm cl}(\leq 1.5)\sim (0.1\mbox{--}2) \times 10^{15}
h^{-1}~{\cal M}_\odot.
\label{eq:25}
\end{equation}
Comparable masses are obtained using the X-ray temperature and
 distribution of
the hot intracluster gas (Hughes 1989; Bahcall and Cen 1993; Lubin and
Bahcall 1993; \S\ref{sec:3}).

The typical (median) blue luminosity of rich clusters (within
1.5$h^{-1}$ Mpc) is
\begin{equation}
L_{\rm cl}(\leq 1.5)\sim 10^{12}h^{-2}~L_\odot
\label{eq:26}
\end{equation}
The approximate range of rich cluster blue luminosities is
\begin{equation}
L_{\rm cl}(\leq 1.5)\sim (0.6\mbox{--}6) \times 10^{12}h^{-2}L_\odot
\label{eq:27}
\end{equation}
The typical mass-to-luminosity ratio of rich clusters is thus 
\begin{equation}
(M/L_{B})_{\rm cl}\sim 300h~({\cal M}_\odot/L_\odot).
\label{eq:28}
\end{equation}
The inferred mass-density in the universe based on cluster dynamics is 
\begin{equation}
\Omega_{\rm dyn}\sim 0.2
\label{eq:29}
\end{equation}
(if mass follows light, $M\propto L$, on scales $\gtorder 1h^{-1}$ Mpc).
$\Omega=1$ corresponds to the critical mass-density needed for a closed
universe and ${\cal M}/L_{B}$ $(\Omega=1)\simeq 1500h$.

\subsection{Cluster Classification}
\label{sec:2.11}

Rich clusters are classified in a sequence ranging from early-
to late-type clusters, or equivalently, from regular to irregular
clusters.  Many cluster properties (shape, concentration,
dominance of brightest galaxy, galactic content, density profile,
and radio and X-ray emission) are correlated with position in
this sequence.  A summary of the sequence and its related
properties is given in Table~\ref{tab:five}.
Some specific classification systems include the Bautz--Morgan (BM) 
 System (Bautz and Morgan 1970), which 
 classifies clusters based on the relative contrast  (dominance
in extent and brightness) of the brightest galaxy to the other
galaxies in the cluster, ranging from type I to III in decreasing
order of dominance; and the Rood-Sastry (RS) system (Rood and Sastry
1971) which classifies clusters based on the distribution of the ten
brightest members (from cD, to binary (B), core (C), line(L), flat
(F), and irregular (I)).

\begin{table}
\caption[]{Cluster Classification and Related Characteristics.
\protect\label{tab:five}}
\begin{tabular*}{5in}{llll}
\noalign{\medskip\hrule\smallskip\hrule\medskip}
 & \multicolumn{1}{c}{Regular (Early)} & \multicolumn{1}{c}{Intermediate} & \multicolumn{1}{c}{Irregular (late)} \\
\multicolumn{1}{c}{Property} & \multicolumn{1}{c}{type clusters} & \multicolumn{1}{c}{clusters} & \multicolumn{1}{c}{type clusters}\\
\noalign{\medskip\hrule\medskip}
Zwicky type &  Compact &  Medium-compact & Open\\
BM type &  I, I--II, II & (II), II--III & (II--III), III\\
RS type &  cD,B,(L,C) & (L),(F),(C) & (F),I\\
Shape symmetry & Symmetrical &  Intermediate &  Irregular shape\\
Central concentration & High & Moderate & Low\\
Galactic content & Elliptical-rich &  Spiral-poor &  Spiral-rich\\
E fraction & 35\% & 20\% & 15\%\\
S0 fraction & 45\% & 50\% & 35\%\\
Sp fraction & 20\% & 30\% & 50\%\\
E:S0:Sp & 3:4:2 & 2:5:3 & 1:2:3\\
Radio emission &$\sim 50\%$ detection 
&$\sim 50\%$ detection&$\sim 25\%$ detection\\
\noalign{\smallskip}
X-ray luminosity & High & Intermediate & Low\\
Fraction of clusters & $\sim 1/3$ & $\sim 1/3$ & $\sim 1/3$\\
Examples &  A401, Coma &  A194 & Virgo, A1228\\
\noalign{\medskip\hrule}
\end{tabular*}
\end{table}

\section{X-Ray Properties of Galaxy Clusters}
\label{sec:3}

\subsection{X-Ray Emission from Clusters}
\label{sec:3.1}

All rich clusters of galaxies produce extended X-ray emission
due to thermal bremsstrahlung radiation from a hot intracluster gas
 (Jones and Forman 1984; Sarazin 1986; David et al. 1993; Edge et al.
1990, 1991; Henry et al. 1991, 1992; Burg et al. 1994).
The cluster X-ray luminosity emitted in the photon energy
band $E_{1}$ to $E_{2}$ by thermal bremsstrahlung from a hot $(T_{\rm x}$
degrees) intracluster gas of uniform electron density $n_{e}$ and a radius
$R_{\rm x}$ is
\begin{equation}
L_{\rm x}\propto n_e^2R_{\rm x}^3T_{\rm x}^{0.5}g
(e^{-E_1/kT_{\rm x}}-e^{-E_2/kT_{\rm x}}).
\label{eq:30}
\end{equation}
The Gaunt factor correction $g$ (of order unity) is a slowly varying
function of temperature and energy.
The bolometric thermal bremsstrahlung luminosity of a cluster
core can be approximated by
\begin{equation}
L_{\rm x}{\rm (core)}\simeq 1.4\times 10^{42}n_{e}({\rm cm}^{-3})^{2}
R_{c}{\rm (kpc)}^{3}kT_{\rm x}{\rm (keV)}^{0.5}h^{-2}~{\rm erg\,s}^{-1}.
\label{eq:31}
\end{equation}

Some of the main properties of the hot intracluster gas
 are summarized below.

\subsection{X-Ray Properties of Clusters}
\label{sec:3.2}

Some of the main properties of the X-ray emission from rich clusters
of galaxies are summarized in Table~\ref{tab:six}.

\begin{table}
\begin{minipage}{5in}
\caption[]{X-Ray Properties of Rich Clusters.\protect\label{tab:six}}
\begin{tabular*}{5in}{ll@{\hspace{-1pt}}c}
\noalign{\medskip\hrule\smallskip\hrule\medskip}
\multicolumn{1}{c}{Property} & \multicolumn{1}{c}{Typical value or range} & Notes\\
\noalign{\medskip\hrule\medskip}
$L_{\rm x}$ (2--10 keV) & $\sim (10^{42.5}\mbox{--}10^{45})h^{-2}$ erg\,s$^{-1}$ & $a$\\
$I_{\rm x}(r)$ & $I_{\rm x}(r)\propto [1+(r/R_{c})^{2}]^{-3\beta +1/2}$ & $b$\\
$\langle\beta\rangle$ & $\sim 0.7$ & $c$\\
$\rho_{\rm gas}(r)$ & $\rho_{\rm gas}(r)\propto
[1+(r/R_{c})^{2}]^{-3\beta /2}$\\
&\ \ \ \ \ \ \ \ \ \  $\propto [1+(r/R_{c})^{2}]^{-1}$ & $d$\\
$kT_{\rm x}$ & $\sim 2$--14 keV & $e$\\
$T_{\rm x}$ & $\sim 2\times 10^{7}$--10$^{8}$ K & $e$\\
$\beta_{\rm spect}={\sigma_r^2\over kT_{\rm x}/\mu m_p}$ & $\sim 1$ & $f$\\
$R_{c}(x)$ & $\sim (0.1$--0.3)$h^{-1}$ Mpc & $g$\\
$n_{e}$ & $\sim 3\times 10^{-3}h^{1/2}$ cm$^{-3}$ & $h$\\
${\cal M}_{\rm gas}$ $(\ltorder 1.5h^{-1}$ Mpc) 
& $\sim 10^{13.5}{\cal M}_\odot$ [range: (10$^{13}$--10$^{14})
{\cal M}_\odot$ h$^{-2.5}]$ & $i$\\
${\cal M}_{\rm gas}/{\cal M}_{\rm cl }$ $(\ltorder
 1.5 h^{-1}$ Mpc) & $\sim 0.07$ (range: 0.03--0.15 h$^{-1.5}$) & $i$\\
Iron abundance & $\sim 0.3$ solar (range: 0.2--0.5) & $j$\\
\noalign{\medskip\hrule\medskip}
\end{tabular*}
\begin{itemize}
\item[$^a$]The X-ray luminosity of clusters (2--10 keV band).  $\langle L_{\rm x}\rangle$
increases with cluster richness and with cluster type (toward
compact, elliptical-rich clusters) (Bahcall 1977a,b; Sarazin 1986; 
Edge et al. 1991; Jones and Forman
1992; David et al. 1993; Burg et al. 1994).
\item[$^b$]X-ray surface brightness distribution, $I_{\rm x}(r)$; $R_{c}$ is the
cluster core radius.
\item[$^c$]Mean $\langle\beta\rangle$ from observations of X-ray brightness profiles
 (Jones and Forman 1984; Sarazin 1986).
\item[$^d$]Implied spatial density profile of the hot gas in the
cluster [from $I_{\rm x}(r)$; isothermal].
\item[$^e$]Range of observed X-ray gas temperature in rich clusters
 (Edge et al. 1990; Henry and Arnaud 1991; Arnaud et al. 1992).
\item[$^f$]$\beta_{\rm spect}$ is the ratio of galaxy to gas 
velocity dispersion:  $\mu$ is
mean molecular weight in amu $(\mu \simeq 0.6)$, $m_{p}$ is mass of the proton,
$\sigma_{r}$ is radial velocity dispersion of 
galaxies in the cluster, and $T_{\rm x}$
is the X-ray temperature of the gas (Lubin and Bahcall 1993).
\item[$^g$]Cluster core radius determined from the X-ray distribution in
the cluster (Jones and Forman 1992).
\item[$^h$]Typical intracluster gas density in rich cluster cores
 (Jones and Forman 1992).
\item[$^i$]Typical mass (and range of masses) of hot gas in rich clusters
and its fraction of the total (virial) cluster mass $({\cal M}_{\rm gas}/
{\cal M}_{\rm cl})$ within $r\ltorder 1.5h^{-1}$ Mpc of the cluster
center (Edge and Stewart 1991; Jones and Forman 1992; White and Fabian
1995; Lubin et al. 1996).
\item[$^j$]Typical iron abundance (and range) of the intracluster gas (in
solar units) (Edge and Stewart 1991; Jones and Forman 1992).
\end{itemize}
\end{minipage}
\end{table}

\subsection{The Intracluster Gas:\ Some Relevant Questions}
\label{sec:3.3}

Some of the fundamental questions about the intracluster gas relate to
its origin, evolution, metal enrichment, hydrodynamical state in the
cluster, and its relation to the distribution of galaxies and mass in
the cluster.  I list below some of the relevant questions and discuss
a selection of these topics in the following subsections.  Some of the
questions posed do not yet have sufficient observational constraints to
suggest a possible solution.

\begin{itemize}
\item What is the hydrodynamical state of the hot gas in clusters?
Is it in approximate hydrostatic equilibrium with the cluster
potential?
\item What is the relation of the intracluster gas to the galaxies and
mass in the cluster?  For example (the subscripts below refer to gas,
galaxies, and mass, respectively):
\begin{description}
\item [---]Density profiles: $\rho_{\rm gas} (r)$ vs. $\rho_{\rm gal}
(r)$ vs. $\rho_m (r)$?
\item [---]Temperature-velocity relation: $T$ vs. $\sigma_{\rm gal}$
vs. $\sigma_m$?
\item [---]Mass:\ $M_{\rm gas}$ vs. $M_{\rm gal}$ vs. $M_{\rm cl}$?
\item [---]Profiles of the above properties:\hfil\break
\hglue.5in\vbox{\hbox{$T(r)$ vs. $\sigma_{\rm gal} (r)$ vs. $\sigma_m
(r)$?}
\hbox{$M_{\rm gas} (r)$ vs. $M_{\rm gal} (r)$ vs. $M_{\rm cl} (r)$?}}
\item [---]Core radii:\ $R_c ({\rm gas})$ vs. $R_c ({\rm gal})$ vs.
$R_c(m)$?
\item [---]Luminosity versus galaxy-type relation:\ $L_x$ vs.  spiral
fraction?
\end{description}
\item What is the origin of the hot intracluster gas?
\item What is the origin of the metal enrichment of the gas?
\item What is the evolution of the intracluster gas?
\end{itemize}

\subsection{The Intracluster Gas:\ Hydrostatic Equilibrium?}
\label{sec:3.4}

The standard model of clusters assumes that
both the gas and the galaxies are in approximate hydrostatic
equilibrium with the binding cluster potential (Bahcall and Sarazin
1977; Forman and Jones 1984; Sarazin 1986; Evrard 1990; Bahcall and
Lubin 1994).  In this model the gas distribution obeys 
\begin{equation}
\frac{dP_{\rm gas}}{dr} = -\frac{GM_{\rm cl} (\leq r) \rho_{\rm gas}}{r^2}
\label{eq:32}
\end{equation}
where $P_{\rm gas}$ and $\rho_{\rm gas}$ are the gas pressure and
density, and $M_{\rm cl}(\leq r)$ is the total cluster binding mass
within a radius $r$.  The cluster mass can thus be represented as
\begin{equation}
M_{\rm cl}(\leq r) = -\frac{kT}{\mu m_p G} \left(\frac{d ln \rho_{\rm
gas} (r)}{d ln r} + \frac{d ln T}{d ln r}\right) r\ ,
\label{eq:33}
\end{equation}
where $T$ is the gas temperature and $\mu m_p$ is the mean particle
mass of the gas.

The galaxies in the cluster respond to the same gravitational field,
and they satisfy
\begin{equation}
M_{\rm cl} (\leq r) = -\frac{\sigma^2_r}{G} \left(\frac{d ln \rho_{\rm gal}
(r)}{d ln r} + \frac{d ln \sigma^2_r}{d ln r} + 2 A\right) r\ ,
\label{eq:34}
\end{equation}
where $\sigma_r$ is the radial velocity dispersion of galaxies in the
cluster, $\rho_{\rm gal} (r)$ is the galaxy density profile, and $A$
represents a possible anisotropy in the galaxy velocity distribution
[$A = 1 - (\sigma_t/\sigma_r)^2$, where $t$ and $r$ represent the
tangential and radial velocity components].

The above two relations yield
\begin{equation}
\beta_{\rm spec} \equiv \frac{\sigma^2_r}{kT/\mu m_p} = \frac{d ln
\rho_{\rm gas} (r)/d ln r + d ln T/d ln r}{d ln \rho_{\rm gal} (r)/d
ln r + d ln \sigma^2_r/d ln r + 2A}\ ,
\label{eq:35}
\end{equation}
where the $\beta_{\rm spec}$ parameter, defined by the left side of
the above relation, can be determined directly from observations of
cluster velocity dispersions and gas temperatures.  The $\beta_{\rm
spec}$ parameter represents the ratio of energy per unit mass in the
galaxies to that in the gas.  Observations of a large sample of
clusters yield a mean best-fit value of $\beta_{\rm spec} \simeq 1 \pm
0.1$ (Lubin and Bahcall 1993; see also \S\ref{sec:3.5}).  This
suggests that, on average, the gas and galaxies follow each other with
comparable energies $(\sigma^2_r \simeq kT/\mu m_p)$.  The observed
mean value $\beta_{\rm spec} \simeq 1 \pm 0.1$ is consistent with the
value of $\beta$ determined from the right-hand side of the $\beta$
relation (referred to as $\beta_{\rm fit}$, and determined from the
gas and galaxy density profile fits).  Using $\rho_{\rm gas}(r)
\propto r^{-2}$ (\S\ref{sec:3.2}) and $\rho_{\rm gal} (r) \propto
r^{-2.4 \pm 0.2}$ (\S\ref{sec:2.6}), one finds $\beta_{\rm fit} \simeq
0.85 \pm 0.1$ (for an isothermal distribution) (Bahcall and Lubin
1994).  The above consistency supports the assumption that the gas is
in an approximate hydrostatic equilibrium with the cluster potential,
and suggests that the galaxies and gas approximately trace each other
in the clusters.

\subsection{The Relation between Gas and Galaxies}
\label{sec:3.5}   

The hot intracluster gas in rich clusters appears to trace reasonably
well the galaxies in the clusters, and---with larger
uncertainty---also the cluster mass.

\begin{figure}
\vglue-2.0in
\plotfiddle{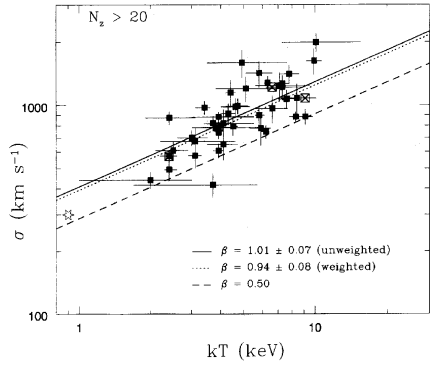}{12cm}{0}{60}{60}{-185}{-90}
\vglue2.0in
\caption[]{Cluster radial velocity dispersion ($\sigma_r$) vs. gas
temperature ($kT$) for 41 clusters (Lubin and Bahcall 1993).  The
best-fit $\beta \equiv \sigma^2_r/(kT/\mu m_p)$ lines are shown by the
solid and dotted curves, with $\beta\simeq 1$.  The $\beta \simeq 0.5$
line previously proposed for a velocity bias in clusters is shown by
the dashed curve;  the velocity bias is inconsistent with the data.}
\label{fig:one}
\end{figure}

\paragraph{Velocity-Temperature relation}

The galaxy velocity dispersion in clusters is well correlated with the
temperature of the intracluster gas; it is observed
(Fig.~\ref{fig:one}) that $\sigma^2_r \simeq kT/\mu m_p$ (Lubin and
Bahcall 1993).  The best-fit $\sigma$--$T$ relation is listed in
\S\ref{sec:3.7}.  The observed correlation indicates that, on average,
the energy per unit mass in the gas and in the galaxies is the same.  
Figure~\ref{fig:one} shows that, unlike previous expectations, the
galaxy velocities (and therefore the implied cluster mass) are not
biased low with respect to the gas (and, by indirect implications,
with respect to the cluster mass; see also \S\ref{sec:3.4}).  Results
from gravitational lensing by clusters also suggest that no
significant velocity bias exists in clusters, and that the gas,
galaxies, and mass provide consistent tracers of the clusters.
Cosmological simulations of clusters (Lubin et al.~1996) produce
$\sigma$--$T$ correlations that match well the data in
Figure~\ref{fig:one}.

\paragraph{Density Profiles}

The gas density profile in clusters follows 
\begin{equation}
\rho_{\rm gas} (r) \simeq \rho_{\rm gas} (o) \left[ 1 +
(r/R_c)^2\right]^{-1}\ ,
\label{eq:36}
\end{equation}
with core radii in the range $R_c \simeq 0.1$--$0.3 h^{-1}$~Mpc
(\S\ref{sec:3.2}).  This implies $\rho_{\rm gas} (r) \propto r^{-2}$
for $R_c < r \ltorder 1.5 h^{-1}~{\rm Mpc}$.

The galaxy density profile in clusters follows approximately (\S\ref{sec:2.6})
\begin{equation}
\rho_{\rm gal} (r) \propto r^{-2.4 \pm 0.2}\ \ \ \ \ \ \ \ R_c < r
\ltorder 1.5 h^{-1}~{\rm Mpc}
\label{eq:37}
\end{equation}
with core radii $R_c \simeq 0.1 - 0.25 h^{-1}~{\rm Mpc}$
(\S\ref{sec:2.7}).

The mass density profile in clusters is less well established, but
initial results from gravitational lensing distortions of background
galaxies by foreground clusters suggest that the mass profile is
consistent with the galaxy density profile (Tyson and Frische 1996).
In the small central core regions of some clusters ($r \ltorder
100$~kpc), the mass distribution may be more compact than the gas or
galaxies, with a small mass core radius of $R_c (m) \ltorder 50
h^{-1}$~kpc.  The results for the overall cluster, however, suggest
that the distributions of gas, galaxies, and mass are similar (with
the gas distribution possibly somewhat more extended than 
the galaxies, as seen by
the mean density slopes above).

\paragraph{Beta-Discrepancy}

The mean $\beta_{\rm spec} \equiv \sigma^2_r/kT/\mu m_p \simeq 1$
result discussed above, combined with the similarity of the gas and
galaxy density profile slopes (that yields $\beta_{\rm fit} \simeq 0.85 \pm
0.1$; \S\ref{sec:3.3}) show that the long claimed $\beta$-discrepancy
for clusters (where $B_{\rm spec} > \beta_{\rm fit}$ was claimed) has
been resolved (Bahcall and Lubin 1994).  The gas and galaxies trace
each other both in their spatial density distribution and in their
energies, as expected for a quasi-hydrostatic equilibrium.

\paragraph{Gas Mass Fraction}

The ratio of the mass of gas in clusters 
to the total virial cluster mass (within $\sim 1.5
h^{-1}$~Mpc) is observed to be in the
range
\begin{equation}
\frac{M_{\rm gas}}{M_{\rm cl}} \simeq 0.03{\rm -}0.15 h^{-1.5}\ ,
\label{eq:38}
\end{equation}
with a median value of
\begin{equation}
\langle \frac{M_{\rm gas}}{M_{\rm cl}} (\ltorder 1.5 h^{-1}~{\rm
Mpc})\rangle_{\rm median} \simeq 0.07 h^{-1.5}
\label{eq:39}
\end{equation}
(Jones and Forman 1992; White et al. 1993; White and Fabian 1995;
Lubin et al. 1996).  The implications of this result, which shows a
high fraction of baryons in clusters, is discussed in \S\ref{sec:4}.

The total gas mass in clusters, $\sim 10^{13}$--$10^{14}
h^{-2.5}M_\odot$, is generally larger than the total mass of the
luminous parts of the galaxies (especially for low values of $h$).
With so much gas mass, it is most likely that a large fraction of the
intracluster gas is of cosmological origin (rather than all the cluster gas
being stripped out of galaxies).  Additional optical-X-ray
correlations of clusters are summarized in \S\ref{sec:3.7}.

\subsection{Metal Abundance in Intracluster Gas}
\label{sec:3.6}

The iron abundance in the intracluster gas is observed to be $\sim
0.3$ solar, with only small variations $(\pm \sim 0.1)$ from cluster
to cluster (e.g., Jones and Forman 1992).  A strong correlation
between the total iron mass in clusters and the total luminosity of
the \hbox{E + SO} cluster galaxies is observed (Jones and Forman
1992).  The metal enrichment of the intracluster gas is likely caused
by gas stripped out of the elliptical galaxy members.  

The iron abundance profile as a function of radius from the cluster
center is generally flat, i.e., a constant abundance at all radii
(except for some poor, low-mass clusters dominated by a single massive
galaxy).  

No evolution is observed in the overall iron abundance of clusters to
$z \sim 0.4$ (Mushotzky 1996).

Recent results using ASCA observations (Mushotzky 1996) of different
element abundances in nearby clusters (O, Ne, Mg, Si, S, Ar, Ca)
suggest a SNII origin for the metals (resulting from early massive
stars) rather than the expected SNIa.  These new results will be
expanded in the near future as additional accurate X-ray data become
available and will provide further clues regarding the origin of the
metallicity of the intracluster gas.

\subsection{X-Ray--Optical Correlations of Cluster Properties }
\label{sec:3.7}

Some observed correlations between X-ray and optical properties
are listed in Table~\ref{tab:seven} (Bahcall 1977a,b; Edge and
Stewart 1991; David et al. 1993; Lubin and Bahcall 1993).

\begin{table}
\begin{minipage}{3.5in}
\caption[]{Correlations Between X-Ray and Optical Properties.$^a$
\protect\label{tab:seven}}
\begin{tabular*}{3.5in}{ll}
\noalign{\medskip\hrule\smallskip\hrule\medskip}
\multicolumn{1}{c}{Properties} & \multicolumn{1}{c}{Correlation}\\
\noalign{\medskip\hrule\medskip}
$\sigma_{r}$-$T$ & $\sigma_{r}$ (km\,s$^{-1})\simeq (332\pm 52)[kT~{(\rm kev)}]^{0.6\pm 0.1}$\\
\\
$T$-$N_{0.5 }$ & $kT$ (keV)$\simeq 0.3N_{0.5_{\phantom{0.5}}}^{0.95\pm 0.18}$\\
\\
$L_{\rm x}$-$N_{0.5}$ & $L_{\rm x}$(bol)$\sim 1.4\times 10^{40}N_{0.5}^{3.16\pm 0.15}h^{-2}$ \\
\\
$L_{x}$-$f_{\rm sp_{\phantom{0.5}}}$ & $L_{\rm x}$(bol)$\simeq 0.6\times 10^{43}f_{\rm sp}^{-2.16\pm 0.11}h^{-2}$\\
\\
$f_{\rm sp}$-$T$ & $f_{\rm sp}\simeq 1.2[kT~({\rm kev})]^{-0.94\pm 0.38}$\\
\\
$T$-$L_{\rm x}$ & $kT$ (keV)$\simeq 0.3[L_{\rm x}{\rm (bol)}h^{2}/10^{40}]^{0.297\pm 0.004}$\\
\noalign{\medskip\hrule\medskip}
\end{tabular*}
\begin{itemize}
\item[$^a$]$\sigma_{r}$ is the galaxy line-of-sight velocity dispersion
in the cluster (km\,s$^{-1}$).
$T$ is the temperature of the intracluster gas $[kT~({\rm keV})]$.
$N_{0.5}$ is the central galaxy density in the cluster [number
of galaxies brighter than $m_{3}+2^{m}$, within $r\leq 0.5h_{50}^{-1}
= 0.25 h^{-1}$ of the cluster center (Bahcall 1977a; Edge and Stewart 1991)].
$L_{\rm x}$(bol) is the bolometric X-ray luminosity of the cluster
(erg\,s$^{-1}$).
$f_{\rm sp}$ is the fraction of spiral galaxies in the cluster
($\ltorder 1.5h^{-1}$ Mpc) (Bahcall 1977b; Edge and Stewart 1991).
Typical uncertainties of the coefficients are $\sim 50\%$ (see
references).
\end{itemize}
\end{minipage}
\end{table}

\subsection{The X-Ray Luminosity Function of Clusters}
\label{sec:3.8}

The X-ray luminosity function of clusters (the number density of
 clusters with X-ray luminosity $L_{\rm x}$ to $L_{\rm x}+dL_{\rm x})$
is approximately (Edge et al. 1990)
\begin{equation}
\Phi_{\rm x}(L_{\rm x})dL_{\rm x}\simeq2.7\times 10^{-7}(L_{\rm x}
/10^{44})^{-1.65}\exp (-L_{\rm x}/8.1\times 10^{44})(dL_{x}
/10^{44})~{\rm Mpc}^{-3}
\label{eq:40}
\end{equation}
for $h = 0.5$, where $L_{\rm x}$ is the 2--10-keV X-ray
luminosity in units of erg\,s$^{-1}$ (for $h=0.5$).
The luminosity function can also be approximated as a power law
 (Edge et al. 1990)
\begin{equation}
\Phi_{\rm x}(L_{\rm x})dL_{\rm x}\simeq 2.2\times 10^{-7}(L_{\rm x}/
10^{44})^{-2.17}(dL_{\rm x}/10^{44})~{\rm Mpc}^{-3}\ (h=0.5).
\label{eq:41}
\end{equation}
The number of X-ray clusters with X-ray luminosity brighter than
$L_{\rm x}$ is approximately
\begin{equation}
n_{c}(>L_{\rm x})\simeq 2\times 10^{-7}(L_{\rm x}/10^{44})^{-1.17}
~{\rm Mpc}^{-3}\ (h=0.5).
\label{eq:42}
\end{equation}
The observed evolution of the X-ray cluster luminosity function suggests
fewer high-luminosity clusters in the past $(z\gtorder 0.5)$ 
 (Edge et al. 1990, Henry et al. 1992). Additional data is required,
however, to confirm and better assess the cluster evolution.

\subsection{Cooling Flows in Clusters}
\label{sec:3.9}

Cooling flows are common at the dense cores of rich clusters;
X-ray images and spectra of $\sim 50\%$ of clusters suggest that the gas
is cooling rapidly at their centers (Sarazin 1986; Fabian 1992).
Typical inferred cooling rates are $\sim 100{\cal M}_\odot/{\rm yr}$.
 The gas cools within $r\ltorder 100h^{-1}$ kpc of the cluster center
(generally centered on the brightest galaxy).
 The cooling flows often show evidence for optical line emission,
blue stars, and in some cases evidence for colder material in HI
or CO emission, or X-ray absorption.

\subsection{The Sunyaev-Zeldovich Effect in Clusters}
\label{sec:3.10}

The Sunyaev--Zeldovich (1972) effect is a perturbation to the
spectrum of the cosmic microwave background radiation as it passes
through the hot dense intracluster gas. It is caused by inverse
Compton scattering of the radiation by the electrons in the cluster
gas.

At the long-wavelength side of the background radiation
spectrum, the hot gas lowers the brightness temperature seen
through the cluster center by the fractional decrement
\begin{equation}
{\delta T\over T}=-2\tau_0{kT_{\rm x} \over m_ec^2},
\label{eq:43}
\end{equation}
where $T=2.73$ K is the microwave radiation temperature, $\tau_{0}$ is the
Thomson scattering optical depth through the cluster
($\tau_{0}=\sigma_{T}\int n_{e}dl$,
where $\sigma_{T}$ is the Thomson scattering cross section and $dl$ is the
distance along the line of sight), $T_{\rm x}$ is the intracluster gas
temperature, and $m_{e}$ is the electron mass.

For typical observed rich cluster parameters of $L_{\rm x}\sim 10^{44}h^{-2}$
erg\,s$^{-1}$, $R_{c}\sim 0.2h^{-1}$ Mpc, and $kT_{\rm x }\simeq 4$ keV, the bremsstrahlung
relation $(L_{\rm x}\propto n_e^2R_c^3T_{\rm x}^{0.5}$, \S\ref{sec:3.1}) implies a
central gas density of $n_{e}\simeq 3 \times 10^{-3}h^{1/2}$ electrons
cm$^{-3}$, thus yielding $\tau_{0}\simeq 3\times 10^{-3}h^{-1/2}$
$[\tau_{0}=0.0064n_{e}({\rm cm}^{-3})R_{c}{\rm (kpc)}]$. Therefore
\begin{equation}
{\delta T \over T}\sim -6\times 10^{-5}h^{-1/2}.
\label{eq:44}
\end{equation}
This temperature decrement remains constant over the cluster core
diameter
\begin{equation}
\theta_{c}\simeq {2H_0R_c \over cz} \simeq {0.5\over z}~{\rm arcmin}
\label{eq:45}
\end{equation}
and decreases at larger separations.

The effect has been detected in observations of some rich, X-ray luminous
clusters (e.g., Coma, A665, A2163, A2218, Cl 0016+16) (Birkinshaw et
al. 1984; Jones et al. 1993; Wilbanks et al. 1994; Herbig et al. 1995).

\section{The Baryon Fraction in Clusters}
\label{sec:4}

Rich clusters of galaxies provide the best laboratory for studying the
baryon fraction (i.e., the ratio of the mass in baryons to the total
mass of the system) on relatively large scales of ${\rm \sim Mpc}$.
The mass of baryons in clusters is composed of {\it at least} two
components: the hot intracluster gas (\S\ref{sec:3}) and the luminous
parts of the galaxies.

The baryon fraction in clusters is therefore
\begin{equation}
\frac{\Omega_b}{\Omega_m} \gtorder \frac{M_{\rm gas} + M_{\rm
stars}}{M_{\rm cl}} \simeq 0.07 h^{-1.5} + 0.05\ ,
\label{eq:46}
\end{equation}
where the first term on the right-hand side represents the gas mass
ratio (\S\ref{sec:3.5}) and the second term corresponds to the stellar
(luminous) contribution.

The baryon density required by big-bang nucleosynthesis is (Walker et
al. 1991)
\begin{equation}
\Omega_b ({\rm BBN}) \simeq 0.015 h^{-2}\ .
\label{eq:47}
\end{equation}
Comparison of the above relations indicates that if $\Omega_m = 1$ then
there are many more baryons observed in clusters than allowed by
nucleosynthesis.  In fact, combining the two relations yields an
$\Omega_m$ value for the mass-density of
\begin{equation}
\Omega_m \ltorder \frac{\Omega_b}{0.07 h^{-1.5} + 0.05} \simeq
\frac{0.015 h^{-2}}{0.07 h^{-1.5} + 0.05} \sim 0.2
\label{eq:48}
\end{equation}
for the observed range of $h \sim 0.5{\rm -}0.8$. Therefore, the baryon
density given by nucleosynthesis and the high baryon content observed
in clusters (mainly in the hot intracluster gas) suggest that
$\Omega_m 
\simeq 0.2$.  This assumes, as expected on this large scale (and as
seen in simulations), that the baryons are not segregated relative to
the dark matter in clusters (White et al. 1993).

\begin{figure}
\vglue-3.0in
\plotfiddle{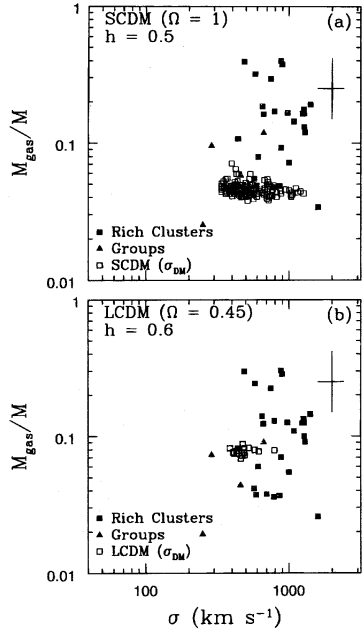}{12cm}{0}{60}{60}{-185}{-90}
\vglue2.0in
\caption[]{Observed and simulated gas mass fraction $(M_{\rm gas}/M)$
vs. line-of-sight velocity dispersion for rich clusters (Lubin et al.
1996).  A typical 1$\sigma$ uncertainty is shown.  The simulated
results (open squares) present the dark matter velocity dispersion;
the galaxy velocity dispersion is lower by $b_v \sim 0.8$ for SCDM and
by $b_v \sim 0.9$ for LCDM.  (a)\ SCDM $(\Omega_m = 1, h = 0.5)$; (b)\
LCDM $(\Omega_m = 0.45,h = 0.6)$.}
\label{fig:two}
\end{figure}

Figure~\ref{fig:two} compares the observed gas mass fraction in
clusters ($\propto$\ baryon fraction) with expectations from
cosmological simulations of $\Omega_m = 1$ and $\Omega_m = 0.45$
cold-dark-matter (CDM) flat models (Lubin et al. 1996).  The results show,
as expected, that the $\Omega_m = 1$ model predicts a much lower gas
mass fraction than observed, by a factor of $\sim 3$.  A low-density
CDM model with $\Omega_m \sim 0.2{\rm -} 0.3$ (in mass)
 best matches the data, as expected
from the general analysis discussed above (see White et al. 1993;
White and Fabian 1995; Lubin et al. 1996).

In summary, the high baryon fraction observed in clusters suggests,
independent of any specific model, that the mass density of the
universe is low, $\Omega_m \sim 0.2{\rm -}0.3$.  This provides a
powerful constraint on high-density $(\Omega_m = 1)$ models; if $\Omega_m
= 1$, a resolution of this baryon problem needs to be found.

\section{Cluster Masses}
\label{sec:5}

The masses of clusters of galaxies within a given radius, $M(\leq r)$,
can be determined from three independent methods:

\begin{description}
\item [a)]Optical:\ galaxy velocity dispersion (and distribution)
assuming hydrostatic equilibrium (\S\ref{sec:3.3});
\item [b)]X-Rays:\ hot gas temperature (and distribution) assuming
hydrostatic equilibrium (\S\ref{sec:3.3});
\item [c)]Lensing:\ gravitational distortions of background galaxies 
(Tyson et al. 1990, 1996; Kaiser and Squires 1993).  This method
determines directly the surface mass overdensity.
\end{description}

The first two methods were discussed in \S\S\ref{sec:2} and 
\ref{sec:3}; the last (and
newest) method is discussed in this book by Narayan.  (See
also Tyson et al. 1990, 1996; Kaiser and Squires 1993; Smail et al.
1995, and references therein).

The galaxy and hot gas methods yield cluster masses that are
consistent with each other as discussed in \S\ref{sec:3} (with
$\sigma^2_r \simeq kT/\mu m_p$).  Gravitational lensing results, which
provide direct information about cluster masses, are available only
for a small number of clusters so far (with data rapidly increasing).
For all but one of the clusters the masses determined from lensing are
consistent, within the uncertainties, with the masses determined from
the galaxies and hot gas methods (see Bahcall 1995 for a summary).
Some differences in masses between the different methods are expected
for individual clusters due to anisotropic velocities, cluster
orientation (yielding larger lensing  surface mass densities for
clusters elongated in the line-of-sight, and vice versa), and
sub-structure in clusters.  On average, however, all three independent
methods yield cluster masses that are consistent with each other.
This triple check on cluster masses provides strong support for the
important cluster mass determinations.

The masses of rich clusters range from $\sim 10^{14}$ to $\sim 10^{15}
h^{-1}~M_\odot$ within $1.5 h^{-1}$~Mpc radius of the cluster center.
When normalized by the cluster luminosity, a robust mass-to-light
ratio is determined for nearby clusters, with only small variations
from cluster to cluster (\S\ref{sec:2.10})
\begin{equation}
\frac{M}{L_B} {\rm (clusters)} \simeq 300 \pm 100 h
\frac{M_\odot}{L_\odot}\ .
\label{eq:49}
\end{equation}
This result is similar to the one obtained from the baryon fraction in
\S\ref{sec:4}. 

If, as desired by theoretical arguments, the universe has critical
mass density $\Omega_m = 1$, than most of the mass in the universe {\it
cannot} be associated with galaxies, groups, and clusters; the mass
distribution in this case would be strongly biased (i.e., mass does
not follow light, with the mass distributed considerably more
diffusely than the light).

\begin{figure}
\vglue-2.0in
\plotfiddle{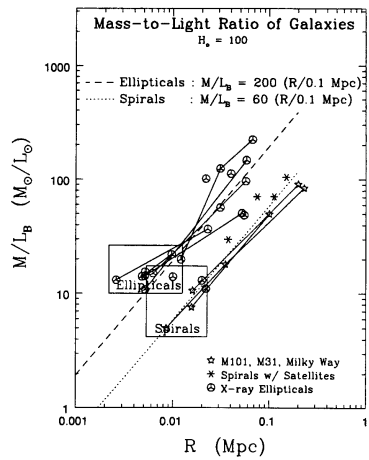}{12cm}{0}{60}{60}{-185}{-90}
\vglue2.0in
\caption[]{Mass-to-light ratio of spiral and elliptical galaxies as a
function of scale (Bahcall, Lubin and Dorman 1995).  The large boxes
indicate the typical $( \sim 1\sigma)$ range of $M/L_B$ for bright
ellipticals and spirals at their luminous (Holmberg) radii.  ($L_B$
refers to {\it total} corrected blue luminosity; see text.)  The
best-fit $M/L_B \propto R$ lines are shown.}
\label{fig:three}
\end{figure}

\section{Where is the Dark Matter?}
\label{sec:6}

\begin{figure}
\vglue-3.0in
\plotfiddle{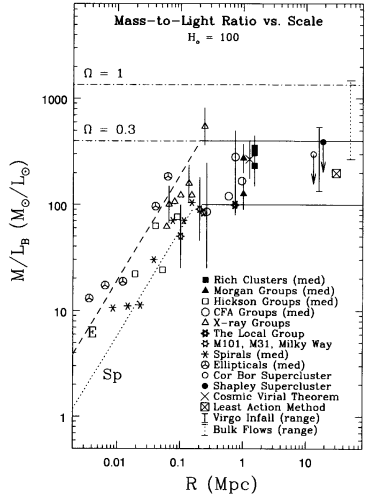}{12cm}{0}{60}{60}{-185}{-90}
\vglue2.0in
\caption[]{Composite mass-to-light ratio of different
systems---galaxies, groups, clusters, and superclusters---as a
function of scale (Bahcall et al. 1995).  The best-fit $M/L_B \propto
R$ lines for spirals and ellipticals (from Fig. \ref{fig:three}) are
shown.  We present median values at different scales for the large
samples of galaxies, groups and clusters, as well as specific values
for some individual galaxies,  X-ray groups, and superclusters.
Typical $1\sigma$ uncertainties and $1\sigma$ scatter around median
values are shown.  Also presented, for comparison, are the $M/L_B$ (or
equivalently $\Omega$) determinations from the cosmic virial theorem,
the least action method, and the {\it range} of various reported
results from the Virgocentric infall and large-scale bulk flows
(assuming mass traces light).  The $M/L_B$ expected for $\Omega = 1$
and $\Omega = 0.3$ are indicated.}
\label{fig:four}
\end{figure}

A recent analysis of the mass-to-light
 ratio of galaxies, groups and clusters by Bahcall, Lubin and Dorman
(1995)  
suggests that while the $M/L$ ratio of galaxies increases with scale
 up to radii of $R \sim 0.1$--$0.2 h^{-1}$~Mpc, due
 to the large dark halos around galaxies (see Fig.~\ref{fig:three};
also Ostriker et al. 1974),  this ratio
 appears to flatten and remain approximately constant for groups and
rich clusters, to scales of $\sim 1.5$~Mpc, and possibly even to the 
larger scales of
superclusters (Fig.~\ref{fig:four}).  The flattening occurs at $M/L_B \simeq
200$--$300 h$, corresponding
to
 $\Omega_m \sim 0.2$.  This observation suggests that most of the dark
 matter is associated with the dark halos of galaxies. Unlike previous
expectations, this result implies that 
 clusters do {\it not} contain a substantial amount of {\it
additional} 
dark matter, other
 than that associated with (or torn-off from) the galaxy halos, and
 the hot intracluster medium.  Bahcall et al. (1995) suggest that the
relatively large $M/L_B$ ratio of clusters $(\sim 300 h)$ results
mainly from a high $M/L_B$ ration of elliptical/SO galaxies.  They
show (Fig.~\ref{fig:three}) that ellipticals have an $M/L_B$ ratio
that is approximately 3 to 4 times larger than typical spirals at the
same radius $[(M/L_B)_s \sim 100 h~{\rm and}~(M/L_B)_e \sim 400 h$
within $r \ltorder 200 h^{-1}~{\rm Kpc}]$.  Since clusters are
dominated by elliptical and SO galaxies, a high $M/L_B$ ratio results.  

Unless the distribution 
 of matter is very different from the distribution of light, with
 large amounts of dark matter in the ``voids" or on very large
 scales, the above results suggest that the mass density in 
the universe may be
 low, $\Omega_m \sim 0.2$ (or $\Omega_m \sim 0.3$ for a small bias
 of $b \sim 1.5$, where the bias factor $b$ relates the overdensity in
galaxies to the overdensity in mass: $b 
\equiv (\Delta\rho/\rho)_{\rm gal}/(\Delta\rho/\rho)_m$).

\section{The Mass Function of Clusters}
\label{sec:7}

\begin{figure}
\vglue-3.0in
\plotfiddle{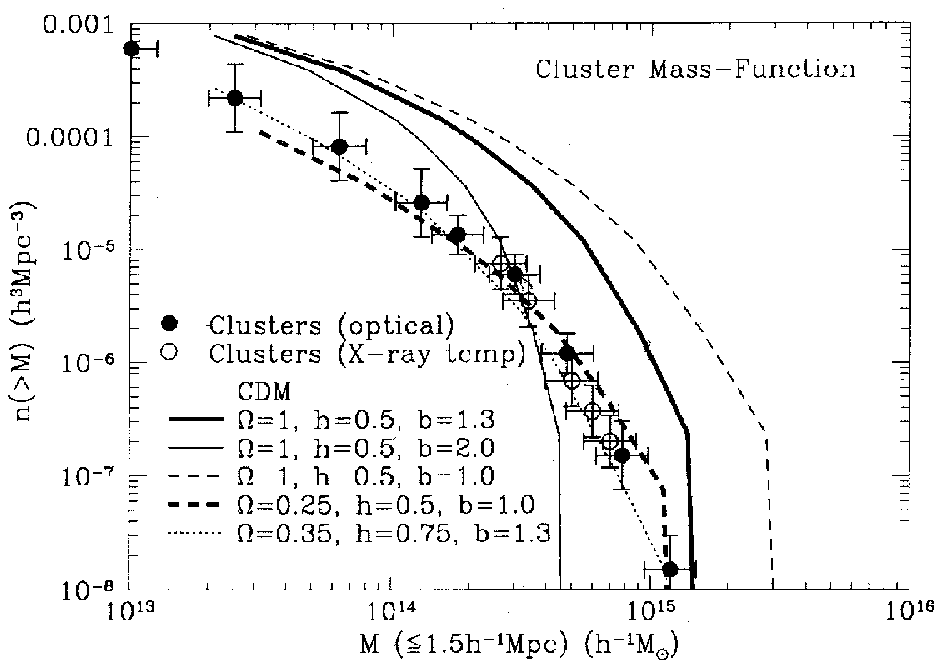}{12cm}{0}{60}{60}{-185}{-90}
\vglue2.0in
\caption[]{Cluster mass functions from observations and from CDM
dimulations (Bahcall and Cen 1992).}
\label{fig:five}
\end{figure}

The observed mass function (MF), $n(>M)$, of clusters of galaxies,
which describes  
the
 number density of clusters above a threshold mass $M$,  
 can be used as a critical test of theories of structure formation in
the universe.  The richest, most massive clusters are thought to 
form from rare high peaks in the
initial
 mass-density fluctuations; poorer clusters and groups form from
smaller, more common
 fluctuations.  
Bahcall and Cen (1993) determined the MF of clusters 
of galaxies using both optical
and
 X-ray observations of clusters.  Their MF is presented in Figure~\ref{fig:five}.  
The function is well fit by the analytic expression
\begin{equation} 
n(>M) = 4 \times 10^{-5} (M/M^*)^{-1} \exp (- M/M^*) h^3~{\rm
Mpc^{-3}}\ ,
\label{eq:50}
\end{equation} 
with $M^* = (1.8 \pm 0.3) \times 10^{14} h^{-1}~ M_\odot$, (where the mass
 $M$ represents the cluster mass within $1.5 h^{-1}$~Mpc radius).
  
The observed cluster mass function is compared in
Figure~\ref{fig:five} with expectations from different
cold-dark-matter cosmologies using large-scale simulations (Bahcall
and Cen 1992).  The comparison shows that the cluster MF is indeed a
powerful discriminant among models.  The standard CDM model $(\Omega_m =
1)$ cannot reproduce the observed MF for any bias parameter; when
normalized to the COBE microwave background fluctuations on large
scales, this model produces too many massive clusters, unseen by the
observations.  A low-density CDM model on the other hand, with
$\Omega_m 
\sim 0.2{\rm -}0.3$ (with or without a cosmological constant),
appears to fit well the observed cluster MF (Fig.~\ref{fig:five}).

\section{Quasar-Cluster Association}
\label{sec:8}

Imaging and spectroscopic data 
 (Yee et al. 1987; Ellingson et al. 1991; Yee et al. 1992) indicate that quasars
are found in environments significantly richer than those of
average galaxies.  The data show a positive association of
quasars with neighboring galaxies.

Optically selected quasars to $z\ltorder 0.7$ exhibit a quasar--galaxy
cross-correlation function amplitude, $A_{qg}$, that is approximately 2.3
times stronger than the galaxy-galaxy correlation amplitude (to
separations 
$r\ltorder 0.25h^{-1}$ Mpc):
\begin{equation}
\langle A_{qg}\rangle \simeq 2.3\langle A_{gg}\rangle \simeq 46.
\label{eq:51}
\end{equation}
This excess correlation suggests that the quasars are typically
located in groups of galaxies with a {\it mean}  richness
\begin{equation}
\langle N_{R}\rangle =n_{g}\int_0^{1.5}A_{qg}r^{-1.8}4\pi
r^2dr\simeq 12~{\rm galaxies}
\label{eq:52}
\end{equation}
(where $n_{g}\simeq 0.015$ Mpc$^{-3}$ is the mean density of galaxies).  The
range of individual group richnesses is, however, wide.

Radio-loud quasars at $z\ltorder 0.5$ are found in similar environments
to those of the optical quasars above.  At $0.5\ltorder z\ltorder 0.7$, the
radio quasars appear to be located in richer environments, with
\begin{equation}
\langle A_{qg}\rangle \simeq 8\langle A _{gg}\rangle \simeq 160
~~~~~~~{\rm (radio~quasars},~ 0.5\ltorder z\ltorder 0.7).
\label{eq:53}
\end{equation}
This cross-correlation amplitude corresponds to a
 {\it mean} environment of rich clusters $(R\sim 0$,
$N_{R}\sim 40)$.  Radio quasars at these redshifts are thus typically
found in rich clusters.

The average galaxy velocity dispersion of the parent clusters
associated with the quasars (Ellingson et al. 1991; Yee et al. 1992) is
\begin{equation}
\sigma_{r}\sim 500~{\rm km\,s}^{-1}.
\label{eq:54}
\end{equation}
The observed auto-correlation function of optically selected
quasars is approximately (Shaver 1988)
\begin{equation}
\xi_{qq}(r,z\sim 0)\simeq 10^{2\pm 0.2}[r({\rm Mpc})]^{-1.8}
\label{eq:55}
\end{equation}
The quasar correlation strength is intermediate between the
correlation of individual galaxies and the correlation of rich clusters.
This correlation strength is consistent with the quasars 
location in groups of the above
mean richness, as would be suggested by the richness-dependent
cluster correlation function (\S\ref{sec:10}).  The quasars may thus
trace the correlation function of their parent clusters (Bahcall and
Chokshi 1991).

Similar results are observed for the association of radio galaxies with
groups and clusters. This association explains the observed 
increase in the strength of the 
radio galaxy correlation function over the general galaxy correlations
 (Bahcall and Chokshi 1992).

\section{Superclusters}
\label{sec:9}

Rich clusters of galaxies are powerful tracers of the large-scale
structure of the universe (Bahcall 1988, Peebles 1993).  I summarize
in the sections below the use of clusters in tracing the large-scale
structure; I include superclusters (\S\ref{sec:9}), statistics of the
cluster correlation function (\S\ref{sec:10}), and peculiar motions on
large scales (\S\ref{sec:11}).

\subsection{Supercluster Properties}
\label{sec:9.1}

Redshift surveys of galaxies reveal that superclusters are very large,
high-density systems of galaxies that are flattened or filamentary in
shape, extending to tens of Mpc. The superclusters appear to surround
large under-dense regions (``voids'') of comparable sizes creating a
``cellular-like'' morphology of the universe on large scales (Gregory
and Thompson 1978; Gregory et al. 1981; Chincarini et al. 1981;
Giovanelli et al. 1986; de-Lapparent et al. 1986; da Costa et al.
1988; Rood 1988; Schectman et al. 1996; Landy et al. 1996).

Large scale superclusters have been identified very effectively by
rich clusters of galaxies (Bahcall and Soneira 1984), like high
mountain peaks tracing mountain chains.  Superclusters are generally
defined as clusters of rich clusters of galaxies above a given spatial
density enhancement $f$.  Here $f \equiv n_c (SC)/n_c$, where $n_c
(SC)$ is the number density of clusters in a supercluster and $n_c$ is
the mean number density of clusters.  The observed superclusters are
large flattened systems, extending to $\sim 150 h^{-1}~{\rm Mpc}$ in
size.  The superclusters typically contain several clusters.  The
high-density superclusters generally surround low-density regions.

A complete catalog of superclusters was constructed by Bahcall and
Soneira (1984) to $z \ltorder 0.08$.  The catalog identifies all
superclusters that have a spatial density enhancement $f \geq 20$.
The mean number density of the Bahcall-Soneira superclusters is $\sim
10^{-6} h^3~{\rm Mpc}^{-3}$, with a mean separation between
superclusters of $\sim 100 h^{-1}~{\rm Mpc}$.  A summary of the main
properties of the superclusters is presented in Table~\ref{tab:eight}.

The superclusters trace well the structure observed in the more
detailed, but smaller, galaxy redshift surveys.

\begin{table}[t]
\begin{minipage}{4in}
\caption[]{Global Properties of Bahcall--Soneira Superclusters.\label{tab:eight}}
\begin{tabular*}{4in}{ll}
\noalign{\medskip\hrule\smallskip\hrule\medskip}
\multicolumn{1}{c}{Property} & \multicolumn{1}{c}{$f=20$ superclusters}\\
\noalign{\medskip\hrule\medskip}
Number density of SCs & $\sim 10^{-6}h^{3}$\,Mpc$^{\rm -3}$\\
Number of clusters per SC & 2--15 clusters\\
Fraction of clusters in SCs & 54\%\\
Size of largest SC & $\sim 150h^{-1}$\,Mpc\\
SC shape &   Flattened\\
Volume of space occupied by SCs & $\sim 3\%$\\
\noalign{\medskip\hrule\medskip}
\end{tabular*}
\end{minipage}
\end{table}

\subsection{Superclusters and Pencil-Beam Surveys}
\label{sec:9.2}

Observations of the redshift distribution of galaxies in narrow $(\sim
40~{\rm arcmin})$ pencil-beam surveys to $z \ltorder 0.3$ (Broadhurst
et al. 1990; hereafter BEKS) reveal a highly clumped and apparently
periodic distribution of galaxies.  The distribution features peaks of
galaxy counts with an apparently regular separation of 128~Mpc, with
few galaxies between the peaks.  What is the origin of this clumpy,
periodic distribution of galaxies?  What does it imply for the nature
of the large-scale structure and the properties discussed above?
Bahcall (1991) investigated these questions observationally, by
comparing the specific galaxy distribution with the distribution of
known superclusters.

Bahcall showed that the observed galaxy clumps originate from the
tails of large superclusters (\S\ref{sec:9.1}).  When the narrow-beams
intersect these superclusters, which have a mean separation of $\sim
100~{\rm Mpc}$, the BEKS galaxy distribution is reproduced.

The redshift distribution of the superclusters is essentially
identical to the galaxy redshift distribution, i.e., it reproduces the
observed peaks in the BEKS survey, for $z \ltorder 0.1$.  This
indicates that the galaxy clumps observed in the pencil-beam survey
originate from these superclusters as the beam crosses the
superclusters' surface.  The main superclusters that contribute to the
clumps were identified.  For example, the first northern clump
originates from the Coma-Hercules supercluster (= the Great-Wall); the
second northern clump is mostly due to the large Corona Borealis
supercluster.

The narrow-beam survey of BEKS is directed toward the north and south
galactic poles.  Some of the Bahcall-Soneira superclusters coincident
with the BEKS peaks are located at projected distances of up to $\sim
50$--100~Mpc from the poles.  This suggests that the high-density
supercluster regions are embedded in still larger halo surfaces, $\sim
100$~Mpc in size, and that these large structures surround large
underdense regions.  The observed number of clumps and their mean
separation are consistent with the number density of superclusters and
their average extent (\S\ref{sec:9.1}).

The narrow widths of the BEKS peaks are consistent with, and imply,
flat superclusters.  From simulations of superclusters and
pencil-beams, Bahcall, Miller, and Udomprasert (1996) find that the
observed peak-widths distribution is consistent with that expected of
randomly placed superclusters with $\ltorder 15$~Mpc width and $\sim
150$~Mpc extent.

The Bahcall-Soneira superclusters may exhibit weak positive
correlations on scales $\sim 100$--150~Mpc (Bahcall and Burgett 1986).
This implies that the superclusters, and thus their related galaxy
clumps, are not randomly distributed but are located in some weakly
correlated network of superclusters and voids, with typical mean
separation of $\sim 100$~Mpc.  This picture is consistent with
statistical analyses of the BEKS distribution as well as with the
observational data of large-scale structure.  The apparent periodicity
in the galaxy distribution suggested by BEKS is expected to be greatly
reduced when pencil-beams in various directions are combined; the
scale reflects the typical {\it mean} separation between large
superclusters, $\sim 100$--$150h^{-1}$~Mpc, but with large variations
at different locations.

\section{The Cluster Correlation Function}
\label{sec:10}

The correlation function of clusters of galaxies 
  efficiently  quantifies the large-scale structure of the
universe.  Clusters are correlated in space more strongly than 
are individual galaxies, by
an order of
 magnitude, and their correlation extends to considerably larger
scales ($\sim 50 h^{-1}$~Mpc).  The cluster correlation strength 
increases with richness ($\propto$ luminosity or mass) of the
 system from single galaxies to the richest clusters
 (Bahcall and Soneira 1983; Bahcall and West 1992).  
 The correlation strength also
increases with
 the mean spatial separation of the clusters
 (Szalay and Schramm 1985; Bahcall and Burgett 1986; Bahcall and West 1992).  
This dependence results in a ``universal"
dimensionless cluster correlation
 function; the cluster dimensionless correlation scale is constant for
all clusters when
 normalized by the mean cluster separation.
  
Empirically, two general relations have been found  (Bahcall and West 1992)
for the correlation function of clusters of galaxies, $\xi_i = A_i
r^{-1.8}$: 
\begin{equation}  
A_i \propto N_i\ ,
\label{eq:56}
\end{equation}
\begin{equation}
A_i \simeq (0.4 d_i)^{1.8}\ ,
\label{eq:57}
\end{equation}
 where $A_i$ is
 the amplitude of the cluster correlation function, 
$N_i$ is the richness 
of
 the galaxy clusters of type $i$ (\S\ref{sec:2.2}), 
and $d_i$ is the mean separation of the
 clusters.  Here $d_i = n_i^{-1/3}$, where $n_i$ is the mean spatial
 number density of clusters of richness $N_i$ (\S\ref{sec:2.3}) in a volume-limited,
richness-limited complete sample.  The first relation, Eq.~(\ref{eq:56}),
states that the amplitude of the cluster correlation function
increases with cluster richness, i.e., rich clusters are more strongly
correlated than poorer clusters.  The second relation, Eq.~(\ref{eq:57}), states
that the amplitude of the cluster correlation function depends on the
mean separation of clusters (or, equivalently, on their number
density); the rarer, large mean separation richer clusters are more
strongly correlated than the more numerous poorer clusters. 
 Eqs.~(\ref{eq:56}) and (\ref{eq:57}) relate to each other 
through the richness function
of clusters, i.e., the number density of clusters as a function of
their richness.  Equation (\ref{eq:57}) 
 describes a universal scale-invariant (dimensionless) correlation
function with  a correlation
 scale $r_{o,i} = A_i^{1/1.8} \simeq 0.4 d_i$ (for $30 \ltorder d_i
\ltorder 90 h^{-1}~{\rm Mpc}$).

There are some conflicting statements in the literature about the
precise values of the correlation amplitude, $A_i$.  Nearly all these
contradictions are caused by not taking account of Eq.~(\ref{eq:56}). 
 When apples are
 compared to oranges, or the clustering of rich clusters is compared
to
 the clustering of poorer clusters, differences are expected and
observed.

\begin{figure}
\vglue-3.0in
\plotfiddle{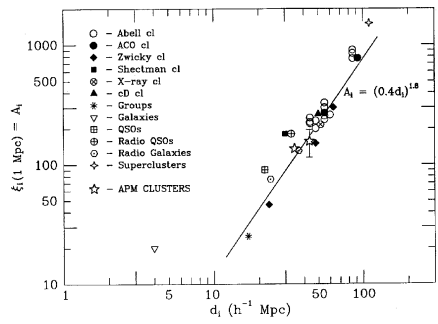}{12cm}{0}{60}{60}{-185}{-90}
\vglue2.0in
\caption[]{The {\it universal} dimensionless cluster correlations: the
dependence of correlation amplitude on mean separation (Bahcall and
 West 1992).  Data points include different samples and catalogs of
clusters and groups, as well as X-ray-selected and cD clusters.
Quasars and radio galaxies, as represented by their parent groups, are
also included.  The APM results are presented; they are consistent
with the expected relation.}
\label{fig:six}
\end{figure}

Figure~\ref{fig:six} clarifies the observational situation.  
The $A_i(d_i)$ relation for groups and clusters of various
richnesses is
 presented in the figure.  The recent automated cluster surveys of APM
 (Dalton et al. 1992) and EDCC (Nichol et al. 1992) are 
consistent with the predictions of Eqs.~(\ref{eq:56}) and (\ref{eq:57}), as is the
correlation
 function of X-ray selected ROSAT clusters of galaxies
 (Romer et al. 1994). 
Bahcall and Cen (1994)
  show that a flux-limited sample of
 X-ray selected clusters will exhibit a correlation scale that is
smaller
 than that of a volume-limited, richness-limited sample of comparable
apparent
 spatial density since the flux-limited sample contains poor groups nearby
and only
 the richest clusters farther away.  Using the richness-dependent
cluster correlations of Eqs.~(\ref{eq:56}) and (\ref{eq:57}), 
Bahcall and Cen (1994) find excellent agreement with the
observed flux-limited
 X-ray cluster correlations of Romer et al. (1994). 

The strong correlation amplitude of galaxy clusters, and the large
scales to which clusters are observed to be positively correlated
$(\sim 50{\rm -}100 h^{-1}~{\rm Mpc})$, complement and quantify the
superclustering of galaxy clusters discussed in \S\ref{sec:9}.
Clusters of galaxies are strongly clustered in superclusters of large
scales (\S\ref{sec:9}), consistent with the strong cluster
correlations to these scales (\S\ref{sec:10}).

\begin{figure}
\vglue-3.0in
\plotfiddle{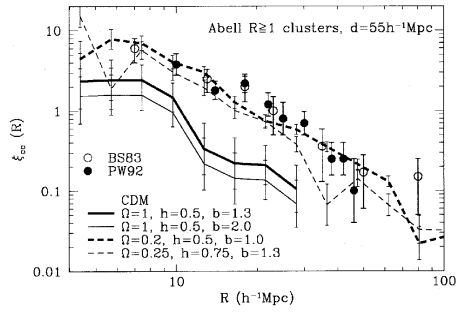}{12cm}{0}{60}{60}{-185}{-90}
\vglue2.0in
\caption[]{Two-point correlation function of Abell $R\geq 1$ clusters,
with mean separation $55 h^{-1}$~Mpc, from observations and the CDM
simulations (Bahcall and Cen 1992).}
\label{fig:seven}
\end{figure}

\begin{figure}
\vglue-3.0in
\plotfiddle{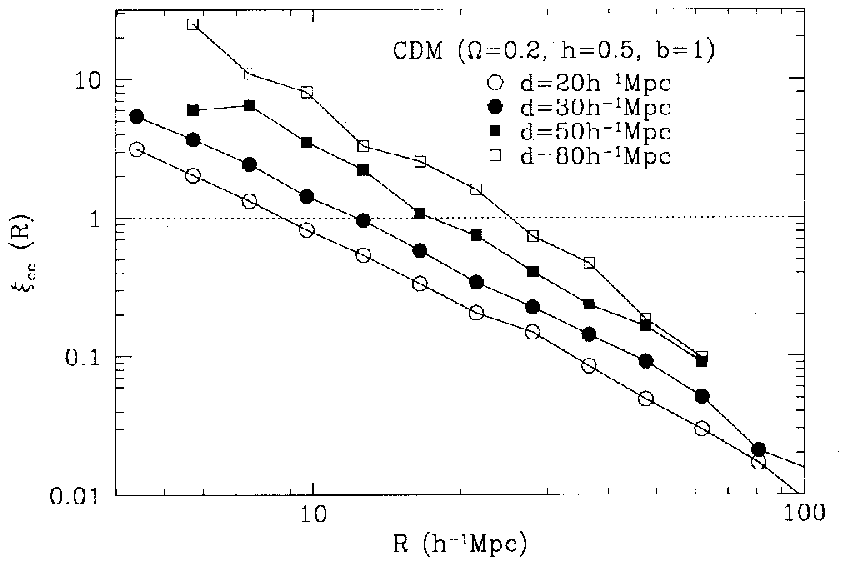}{12cm}{0}{60}{60}{-185}{-90}
\vglue2.0in
\caption[]{Model dependence of the cluster correlation function on
mean separation $d$ (CDM simulation: $\Omega = 0.2, h = 0.5, b = 1$)
(from Bahcall and Cen 1992).}
\label{fig:eight}
\end{figure}

\begin{figure}
\vglue-3.0in
\plotfiddle{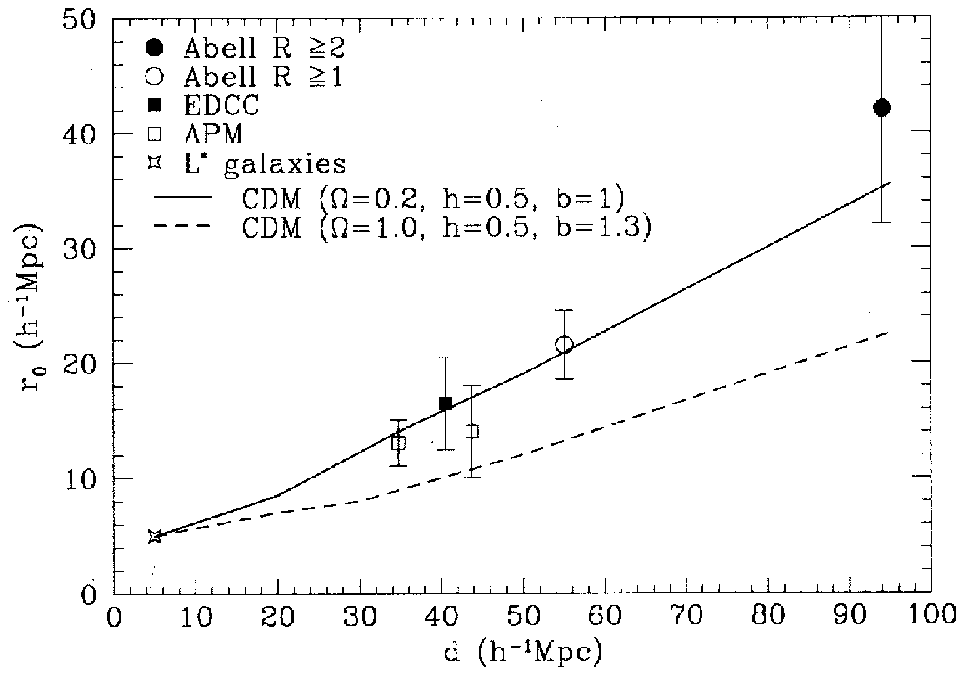}{12cm}{0}{60}{60}{-185}{-90}
\vglue2.0in
\caption[]{Correlation length as a function of cluster separation,
from both observations and simulations (Bahcall and Cen 1992).}
\label{fig:nine}
\end{figure}

This fundamental observed property of clusters of galaxies---the
cluster correlation function---can be used to place strong constraints
on cosmological models and the density parameter $\Omega_m$ by
comparison with model expectations.  Bahcall and Cen (1992) contrasted
these cluster observations with standard and nonstandard CDM models
using large N-body simulations ($400 h^{-1}$ box, $10^{7.2}$
particles).  They find that none of the standard $\Omega_m = 1$ CDM
models can fit consistently the strong cluster correlations.  A
low-density ($\Omega_m \sim 0.2{\rm -}0.3$) CDM-type model (with or
without a cosmological constant), however, provides a good fit to the
cluster correlations (see Figs. \ref{fig:seven}--\ref{fig:nine}) as
well as to the observed cluster mass-function (\S\ref{sec:7}, Fig.
\ref{fig:five}).  This is the first CDM-type model that is consistent
with the high amplitude and large extent of the correlation function
of the Abell, APM, and EDCC clusters.  Such low-density models are
also consistent with other observables as discussed in this paper.
The $\Omega_m$ constraints of these cluster results are model
dependent; a mixed hot + cold dark matter model, for example, with
$\Omega_m = 1$, is also consistent with these cluster data (see
Primack's chapter in this book).

The CDM results for clusters corresponding to the
 rich Abell clusters $({\rm richness~class}\ R \geq 1)$ with $d = 55 h^{-1}$~Mpc are
 presented in Figure~\ref{fig:seven} together with the observed correlations 
 (Bahcall and Soneira 1983; Peacock and West 1992).  The results indicate that the
standard
$\Omega_m = 1$ CDM models are inconsistent with the observations; they cannot
provide
 either the strong amplitude or the large scales ($\gtorder 50 h^{-1}$~Mpc) to 
which the cluster correlations are observed.  Similar results are
found for
 the APM and EDCC clusters.

The low-density, low-bias model is 
 consistent with the data;  it 
reproduces both the strong
 amplitude and the large scale to which the cluster correlations are
detected.  


The dependence of
 the observed cluster correlation on $d$ was also tested in the
simulations.  
The results are shown in Figure~\ref{fig:eight} for the
 low-density model.  The dependence of correlation amplitude on mean
separation is
 clearly seen in the simulations.  To compare this result directly
with observations, I plot in Figure~\ref{fig:nine} the dependence of the 
correlation scale, $r_o$, on $d$ for both the simulations and 
the observations.  The low-density
 model agrees well with the observations, yielding $r_o \approx 0.4 d$, as
observed.  The $\Omega_m = 1$ model, while also showing an increase of $r_o$ with
 $d$, yields considerably smaller correlation scales and a much slower
increase of
 $r_o (d)$.

What causes the observed dependence on cluster richness 
[Eqs.~(\ref{eq:56}--\ref{eq:57}]?
The dependence, seen
both
 in the observations and in the simulations, is most likely caused by
 the statistics of rare peak events, which Kaiser (1984)  
suggested as an explanation of the observed strong increase of correlation amplitude from
galaxies to
 rich clusters.  The correlation function of rare peaks in a Gaussian
field increases with their selection threshold.  Since more massive
clusters correspond to a
 higher threshold, implying rarer events and thus larger mean
separation, Eq.~(\ref{eq:57}) results.  A 
 fractal distribution of galaxies and clusters would also produce 
Eq.~(\ref{eq:57}) (e.g., Szalay and Schramm 1985).

\section{Peculiar Motions of Clusters}
\label{sec:11}

 How is the mass distributed in the universe?  Does it follow, on the
average,  the light
distribution? To address this important question, peculiar motions on large
scales are studied
 in order to directly trace the mass distribution.  It
 is believed that the peculiar motions (motions relative to a pure
Hubble
 expansion) are caused by the growth of cosmic
structures due to gravity.  
A comparison
 of the mass-density distribution, as reconstructed from peculiar
velocity data, with
 the light distribution (i.e., galaxies) provides information on how
well
 the mass traces light (see chapter by Dekel, and 1994). The basic underlying
relation between peculiar velocity and density is given by 
\begin{equation}
\vec\nabla \cdot \vec v = - \Omega_m^{0.6} \delta_m = -\Omega_m^{0.6}
\delta_g/b
\label{eq:59}
\end{equation}
where $\delta_m \equiv (\Delta \rho/\rho)_m$ is the mass overdensity,
$\delta_g$ is the galaxy overdensity, and $b \equiv \delta_g/\delta_m$
is the bias parameter discussed in \S\ref{sec:6}.
 A formal 
analysis yields a measure of the
 parameter $\beta \equiv \Omega_m^{0.6}/b$. 
Other   methods that place
constraints on $\beta$
 include the anisotropy in the galaxy distribution in the redshift
direction due to peculiar motions (see Strauss and Willick 1995 for a
review).

Measuring peculiar motions is difficult.  The motions are
usually inferred with the aid of measured distances to galaxies or clusters
that are obtained  using some (moderately-reliable) 
distance-indicators (such as
 the Tully-Fisher or $D_n -\sigma$ relations), and 
the measured galaxy redshift.  The peculiar velocity $v_p$ is 
 then determined from the difference between the measured redshift
velocity, $cz$, and
 the measured Hubble velocity, $v_H$, of the system (the latter obtained from the
distance-indicator):  $v_p = cz - v_H$.

\begin{figure}
\vglue-3.0in
\plotfiddle{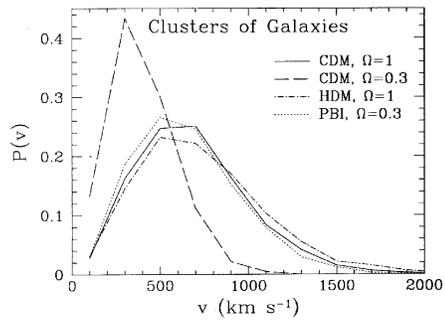}{12cm}{0}{60}{60}{-185}{-90}
\vglue2.0in
\caption[]{Differential three-dimensional peculiar velocity
distribution of rich clusters of galaxies for four cosmological models
(Bahcall, Gramann and Cen 1994).}
\label{fig:ten}
\end{figure}

A summary of all  measurements of $\beta$ made so far is 
presented in Strauss and Willick (1995).  The dispersion
 in the current measurements of $\beta$ is very large; the various
determinations
 range from $\beta \sim 0.4$ to $\sim 1$, implying, for $b \simeq 1, \Omega_m
 \sim 0.2$ to $\sim 1$.  No strong conclusion can therefore be reached at present
 regarding the values of $\beta$ or $\Omega_m$.  The larger and more accurate
 surveys currently underway, including high precision velocity
measurements, will likely lead to the
 determination of $\beta$ and possibly its decomposition into
$\Omega_m$ and $b$ (e.g., Cole et al. 1994).

\begin{figure}
\vglue-3.0in
\plotfiddle{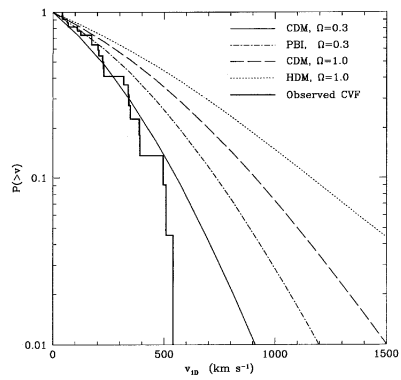}{12cm}{0}{60}{60}{-185}{-90}
\vglue2.0in
\caption[]{Observed vs. model cluster peculiar velocity functions
(from Bahcall and Oh 1996).  The Giovanelli and Haynes (1996) data are
compared with model expectations convolved with the observational
errors.  Note the absence of a high velocity tail in the observed
cluster velocity function.}
\label{fig:eleven}
\end{figure}

Clusters of galaxies can also serve as
 efficient tracers of the large-scale peculiar velocity field in the
universe (Bahcall, Gramann and Cen 1994).  
Measurements of cluster peculiar velocities are
 likely to be more
 accurate than measurements of individual 
galaxies, since cluster distances can be
determined by averaging
 a large number of cluster members as well as by using different
distance indicators.  Using large-scale cosmological simulations, 
Bahcall et al. (1994)
find that clusters
 move reasonably fast in all the cosmological models studied, tracing well the
underlying matter
 velocity field on large scales.  The clusters exhibit a Maxwellian
distribution of
 peculiar velocities as expected from Gaussian initial density
fluctuations.  The model cluster 3-D velocity distribution, 
presented in Figure~\ref{fig:ten}, typically peaks at $v \sim
600~{\rm km~ s^{-1}}$ and extends to high cluster velocities of 
$\sim 2000~{\rm km~ s^{-1}}$.  The 
low-density CDM model exhibits lower velocities
(Fig.~\ref{fig:ten}).  Approximately 10\% of all model rich clusters (1\% for low-density
 CDM) move with $v \gtorder  10^3~{\rm km~ s^{-1}}$.  A comparison of model
 expectation with recent, well calibrated cluster velocity data
(Giovanelli et al. 1996) is
presented in Figure~\ref{fig:eleven} (Bahcall and Oh 1996).  
The comparison between models and observations suggests that the
cluster velocity data is consistent with a low-density CDM model, and
is inconsistent with a standard $\Omega_m = 1$ CDM model, since no high
velocity clusters are observed.

\begin{figure}
\vglue-3.0in
\plotfiddle{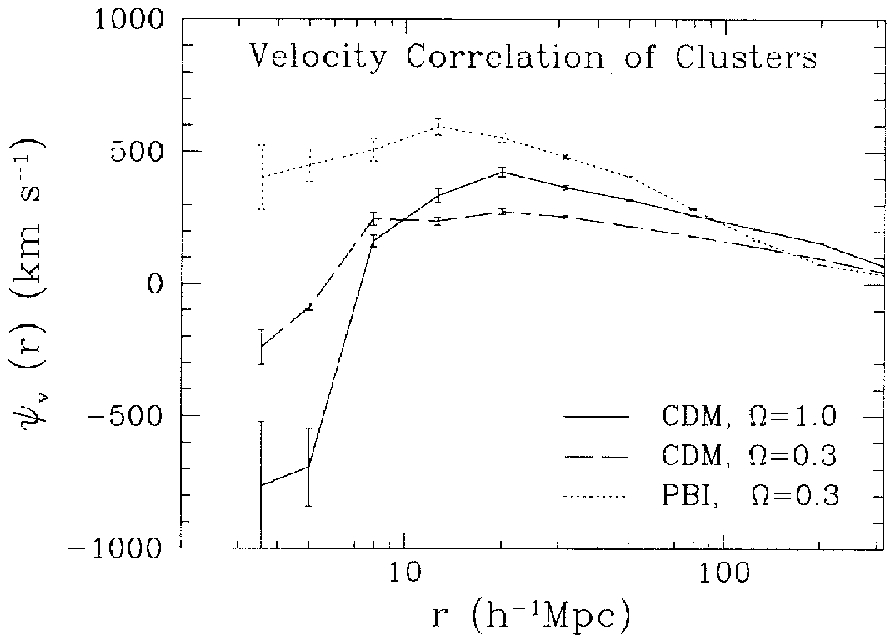}{12cm}{0}{60}{60}{-185}{-90}
\vglue2.0in
\caption[]{Velocity correlation function of rich $(R \geq 1)$ clusters
of galaxies for three models.  Error bars indicate the $1\sigma$
statistical uncertainties (from Cen et al. 1994).}
\label{fig:twelve}
\end{figure}

Cen, Bahcall and Gramann (1994) determined the
expected velocity correlation function
 of clusters in different cosmologies.  They find that close cluster
pairs, with
 separations $r \ltorder 10 h^{-1}$~Mpc, exhibit strong attractive motions; the
pairwise velocities
 depend sensitively on the model.  The mean pairwise attractive
cluster velocities on
 $5 h^{-1}$~Mpc scale ranges from $\sim 1700~{\rm km~ s^{-1}}$ for 
$\Omega_m = 1$ CDM to $\sim 700~{\rm km~ s^{-1}}$ for $\Omega_m = 0.3$
CDM.  
The cluster velocity correlation function,
presented in Figure~\ref{fig:twelve}, is negative 
on small scales---indicating large attractive
velocities, and is
 positive on large scales, to $\sim 200 h^{-1}$~Mpc---indicating significant
bulk motions
 in the models.  None of the models reproduce the very large
 bulk flow of clusters on $150 h^{-1}$~Mpc scale, $v \simeq 689 \pm 178~
{\rm km~ s^{-1}}$, recently reported by Lauer and Postman (1994).  The bulk
 flow expected on this large scale is generally  $\ltorder 200~{\rm km~ s^{-1}}$
 for all the models studied ($\Omega_m = 1$ and 
$\Omega_m ~\simeq 0.3$ CDM, and PBI).

\section{Some Unsolved Problems}
\label{sec:12}

Considerable progress has been made over the last two decades in the
study of clusters and superclusters of galaxies, as described in these
lectures.  However, many problems still remain open.  I highlight some
of the unsolved problems in this field that are likely to be solved in
the coming decade.  Currently planned large redshift surveys of
galaxies and clusters such as the Sloan Digital Sky Survey and the 2dF
survey, deep optical and X-ray surveys using HST, Keck, ROSAT, ASCA,
and AXAF, among other, should allow a considerable increase in our
understanding of the nature and evolution of these fundamental
systems.  At the same time, state of the art cosmological simulations
to be available in the next decade (e.g., Ostriker, this book)
should greatly enhance our ability to compare the observations with
detailed expectations from various cosmologies and hopefully narrow
down the correct cosmological model of our universe.

Here is a partial list of some of the interesting unsolved problems in
the field of clusters and superclusters of galaxies.

\paragraph{Clusters of Galaxies}

\begin{itemize}
\item What is the mass distribution and its extent in clusters of
galaxies?  Using gravitational lensing distortions, one can determine
the mass density profile, $\rho_m (r)$, and total cluster mass,
$M(r)$, of clusters and compare it with the distribution of galaxies
and gas for a large sample of clusters.
\item Does mass follow light on these scales?  If not, what is the
bias factor as a function of scale, $b(r)$?
\item What is the implied density parameter from clusters,
$\Omega_m(r)$?
\item What is the accurate baryon-fraction in clusters and groups of
galaxies, as a function of scale, $\Omega_b/\Omega (r)$?
\item What is the origin of the hot intracluster gas and its
metallicity?
\item What is the evolution of clusters in the optical and in X-rays?
\item What are the cosmological implications from studies of
clusters?
\end{itemize}

\paragraph{Superclusters}

\begin{itemize}
\item What is, quantitatively, the morphology of superclusters and
large-scale structure (superclusters, filaments, and void network)?
\item What is the dependence of the superclustering properties on
galaxy luminosity, surface brightness, type (E, S), and system
(galaxies versus clusters)?
\item What are the peculiar motions in superclusters and on large
scales?
\item What is the mass, and mass distribution, in superclusters and on
large scales?  Does mass follow light?
\item What is in the ``voids''?
\item What is $\Omega_m$ on large scales?
\item What is the baryon fraction in superclusters?
\item What is the time evolution of superclusters?
\item What are the constraints made by the observed superclusters and
large-scale structure on cosmology and galaxy formation models?
\end{itemize}

I expect that many of these questions will be addressed and possibly
solved in the coming decade.

\section*{Acknowledgments}

I thank the organizers of the Jerusalem Winter School 1995, J. P.
Ostriker and A. Dekel, for an outstanding, productive, and fun  school.  The
work by N. Bahcall and collaborators is supported by NSF grant AST93-15368.


\begin{thebibliography}{100}
\bibitem{}
Abell, G. O. 1958, ApJS, 3, 211
\bibitem{}
Arnaud, M., Hughes, J. P., Forman, W., Jones, C., Lachieze-Rey, M.,
Yamashita, K., \& Hatusukade, I. 1992, ApJ, 390, 345
\bibitem{}
Bahcall, J. N., \& Sarazin, C. 1977, ApJ, 213, L99
\bibitem{}
Bahcall, N. A. 1975, ApJ, 198, 249
\bibitem{}
Bahcall, N. A. 1977, ARA\&A, 15, 505
\bibitem{}
Bahcall, N. A. 1977a, ApJ, 217, L77
\bibitem{}
Bahcall, N. A. 1977b, ApJ, 218, L93
\bibitem{}
Bahcall, N. A. 1981, ApJ, 247, 787
\bibitem{}
Bahcall, N. A. 1988, ARA\&A,  26, 631
\bibitem{}
Bahcall, N. A. 1991, ApJ, 376, 43
\bibitem{}
Bahcall, N. A. 1995, in Dark Matter, AIP Conf. Proceedings 336, ed. S.
Holt and C. Bennet (New York: AIP), 201
\bibitem{}
Bahcall, N. A. 1996, in Astrophysical Quantities, ed. A. Cox (New
York: AIP)
\bibitem{}
Bahcall, N. A., \& Burgett, W. 1986, ApJL, 300, L35
\bibitem{}
Bahcall, N. A., \& Cen, R. 1992, ApJ, 398, L81
\bibitem{}
Bahcall, N. A., \& Cen, R. 1993, ApJ, 407, L49
\bibitem{}
Bahcall, N.~A., \& Cen, R.~Y. 1994, ApJ, 426, L15
\bibitem{}
Bahcall, N. A., \& Chokshi, A. 1991, ApJ, 380, L9
\bibitem{}
Bahcall, N. A., \& Chokshi, A. 1992, ApJL, 385, L33
\bibitem{}
Bahcall, N.~A., Gramann, M., \& Cen, R. 1994, ApJ, 436, 23.
\bibitem{}
Bahcall, N. A., \& Lubin, L. M. 1994, ApJ, 426, 513
\bibitem{}
Bahcall, N. A., Lubin, L M., \& Dorman, V. 1995, ApJ, 447, L81
\bibitem{}
Bahcall, N. A., Miller, N., \& Udomprasert, P. 1996, in preparation 
\bibitem{}
Bahcall, N. A., \& Oh, P. 1996, ApJ, 462, L43
\bibitem{}
Bahcall, N. A., \& Soneira, R. M. 1983, ApJ, 270, 20
\bibitem{}
Bahcall, N. A,. \& Soneira, R.M., 1984, ApJ, 277, 27 
\bibitem{}
Bahcall, N. A., \& West, M. 1992, ApJL, 392, 419
\bibitem{}
Bautz, L. P., \& Morgan, W. W. 1970, ApJ, 162, L149
\bibitem{}
Birkinshaw, M., Gull, S. F., \& Hardebeck, H. E. 1984, Nature,
 309, 34
\bibitem{}
Broadhurst, T. J., Ellis, R., Koo, D., \& Szalay, A. 1990, Nature,
343, 726
\bibitem{}
Burg, R., Giacconi, R., Forman, W., Jones, C. 1994, ApJ, 422, 37
\bibitem{}
Cen, R., Bahcall, N.~A., \& Gramann, M. 1994, ApJ, 437, L51
\bibitem{}
Chincarini, G., Rood, H.~J., \& Thompson, L.~A. 1981, ApJ, 249, L47
\bibitem{}
Cole, S., Fisher, K.~B., \& Weinberg, D.~H. 1994, MNRAS, 267, 785
\bibitem{}
da Costa, L.~N., et al. 1988, ApJ, 327, 544
\bibitem{}
Dalton, G. B., Efstathiou, G., Maddox, S. J., \& Sutherland, W.
1992, ApJ, 390, L1
\bibitem{}
David, L. P., Slyz, A., Jones, C., Forman, W., Vrtilek, S., \&
Arnaud, K. 1993, ApJ, 412, 479
\bibitem{}
de Lapparent, V., Geller, M., \& Huchra, J. 1986, ApJ, 302, L1
\bibitem{}
Dekel, A. 1994, ARA\&A, 32, 371
\bibitem{}
Dressler, A. 1978, Ap.J, 226, 55
\bibitem{}
Dressler, A. 1980, ApJ, 236, 351
\bibitem{}
Dressler, A. 1984, ARA\&A, 22, 185
\bibitem{}
Edge, A., \& Stewart, G. C., 1991, MNRAS, 252, 428
\bibitem{}
Edge, A., Stewart, G. C., Fabian, A. C., \& Arnaud, K. A. 1990,
MNRAS, 245, 559
\bibitem{}
Ellingson, E., Yee, H. K. C., \& Green, R. F. 1991, ApJ, 371, 45
\bibitem{}
Evrard, A. E. 1990, ApJ, 363, 349
\bibitem{}
Fabian, A. C. 1992, in Clusters and Superclusters of Galaxies,
NATO ASI Series No. 366, edited by A. C. Fabian (Dordrecht: Kluwer Academic), p.151
\bibitem{}
Geller, M. J. 1990, in Clusters of Galaxies, STScI Symposium
No. 4, ed. W. R. Oegerle et al. (Cambridge: Cambridge Univerity
Press), p. 25
\bibitem{}
Giovanelli, R., \& Haynes, M. 1996, in preparation
\bibitem{}
Giovanelli, R., Haynes, M., \& Chincarini, G. 1986, ApJ, 300, 77
\bibitem{}
Gregory, S.~A., \& Thompson, L.~A. 1978, ApJ, 222, 784
\bibitem{}
Gregory, S.~A., Thompson, L.~A., \& 
Tifft, W. 1981, ApJ, 243, 411
\bibitem{}
Henry, J. P., \& Arnaud, K. A. 1991, ApJ, 372, 410
\bibitem{}
Henry, J. P., Gioia, I. M., Maccacaro, T., Morris, S. L., Stocke, J.
T., \& Walter, A. 1992, ApJ, 386, 408
\bibitem{}
Herbig, T., Lawrence, C. R., Readhead, A. C. S., \& Gulkis, S.
1995, ApJL, 449, L1
\bibitem{}
Hoessel, J. G., Gunn, J. E., \& Thuan, T. X. 1980, ApJ, 241, 486
\bibitem{}
Hughes, J. P. 1989, ApJ, 337, 21
\bibitem{}
Jones, C., \& Forman, W. 1984, ApJ, 276, 38
\bibitem{}
Jones, C., \& Forman, W. 1992, in Clusters and Superclusters of
Galaxies, NATO ASI Series, No. 366, ed. A. C. Fabian
(Dordrecht: Kluwer Academic), p. 49
\bibitem{}
Jones, M., Saunders, R., Alexander, P., Birkinshaw, M., \&
Dillon, N. 1993, Nature, 365,  320
\bibitem{}
Kaiser, N., \& Squires, G. 1993, ApJ, 404, 441
\bibitem{}
King, I. 1972, ApJL, 174, L123
\bibitem{}
Landy, S., Shectman, S., Lin, H., Kirshner, R., Oemler, A., \& Tucker,
A. 1996, ApJ, 456, L1
\bibitem{}
Lauer, T., \& Postman, M. 1994, ApJ., 425, 418
\bibitem{}
Lilje, P. B., \& Efstathiou, G. 1988, MNRAS, 231, 635
\bibitem{}
Lubin, L., \& Bahcall, N. A. 1993, ApJL, 415, L17
\bibitem{}
Lubin, L., Cen, R., Bahcall, N. A., \& Ostriker, J. P. 1996, ApJ, 460, 10
\bibitem{}
Lumsden, S. L., Nichol, R. C., Collins, C. A., \& Guzzo, L.
1992, MNRAS, 258, 1
\bibitem{}
Mushotzky, R. 1996, preprint 
\bibitem{} 
Nichol, R., Collins, C. A., Guzzo, L., \& Lumsden, S. L. 1992,
MNRAS,255, 21p
\bibitem{}
Oemler, A. 1974, ApJ, 194, 1
\bibitem{}
Oort, J. 1983, ARA\&A, 21, 373
\bibitem{}
Ostriker, J. P., Peebles, P. J. E., \& Yahil, A. 1974, ApJ, 193, L1
\bibitem{}
Peacock, J., \& West, M. 1992, MNRAS, 259, 494
\bibitem{}
Peebles, P. J. E. 1980, The Large Scale Structure of the
Universe (Princeton: Princeton University Press)
\bibitem{}
Peebles, P. J. E. 1993, Principles of Physical Cosmology
(Princeton: Princeton University Press)
\bibitem{}
Postman, M., \& Geller, M. 1984, ApJ, 281, 95
\bibitem{}
Romer, A.K., Collins, C., B\"ohringer, H., 
Cruddace, R., Ebeling, H.,
MacGillivray, H., \& Voges, W. 1994, Nature, 372, 75
\bibitem{}
Rood, H. J. 1988, ARA\&A, 26, 245
\bibitem{}
Rood, H. J., \& Sastry, G. N. 1971, PASP, 83, 313
\bibitem{}
Sarazin, C. L. 1986, Rev. Mod. Phys., 58, 1
\bibitem{}
Schechter, P. L. 1976, ApJ, 203, 297
\bibitem{}
Shaver, P. 1988, in Large-Scale Structure of the Universe,
IAU Symposium No. 130, ed. J. Audouze et al. (Dordrecht: Reidel), 359
\bibitem{}
Shectman, S., Landy, S., Oemler, A., Tucker, A., Lin, H., \& Kirshner,
R. 1996, ApJ, in press (preprint astro-ph/9604167)
\bibitem{}
Smail, J., Ellis, R., Fitchett, M., \& Edge, A. 1995, MNRAS, 273, 277
\bibitem{}
Strauss, M., \& Willick, J. 1995, Phys. Reports, 261, 271
\bibitem{}
Struble, M., \& Rood, H. 1991, ApJS, 77, 363
\bibitem{}
Sunyaev, R. A., \& Zeldovich, Ya. B., 1972, Comments Astrophys. Space
Phys., 4, 173
\bibitem{}
Szalay, A., \& Schramm, D. N. 1985, Nature, 314, 718
\bibitem{}
Tyson, J. A., \& Fischer, P. 1996, ApJL, in press
\bibitem{}
Tyson, J. A., Valdes, F., \& Wenk, R. A. 1990, ApJ, 349, Ll
\bibitem{}
Walker, T. P., et al. 1991, ApJ, 376, 51
\bibitem{}
White, D., \& Fabian, A. 1995, MNRAS, 273, 72
\bibitem{}
White, S. D. M., Navaro, J. F., Evrard, A. E., \& Frenk, C. S. 1993
Nature, 366, 429
\bibitem{}
Wilbanks, T. M., Ade, P. A. R., Fischer, M. L., Holzapfel, W. L.,
\& Lange, A. 1994, ApJ, 427, L75
\bibitem{}
Yee, H. K. C. 1992, in Clusters and Superclusters of Galaxies,
NATO ASI Series No. 366, ed.  A. C. Fabian (Dordrecht: Kluwer Academic), 293
\bibitem{}
Yee, H. K. C., \& Green, R. F. 1987, ApJ, 319, 28
\bibitem{}
Zwicky, F. 1957, Morphological Astronomy (Berlin: Springer)

\end{thebibliography}
\end{document}